\documentclass[floatfix,reprint,amsmath,amssymb,aps,prx,longbibliography,bibnotes]{revtex4-2}

\usepackage{graphicx}
\usepackage{bm}
\usepackage[hidelinks]{hyperref}
\usepackage{amsmath}
\usepackage[capitalize]{cleveref}
\usepackage{physics}
\usepackage{xcolor}
\usepackage{xspace}

\DeclareMathOperator{\sign}{sign}
\newcommand{\PTsymm}{$\mathcal{PT}$ symmetry\xspace}
\newcommand{\Usymm}{$U(1)$ symmetry\xspace}
\newcommand{\PTtrans}{$\mathcal{PT}$ transformation\xspace}
\newcommand{\pump}{\gamma_+}
\newcommand{\cavdec}{\kappa}
\newcommand{\meas}{\mathrm{m}}
\newcommand{\amod}{a_+}

\begin{document}
\date{December 3, 2024}

\title{Nonreciprocal Synchronization of Active Quantum Spins}

\author{Tobias Nadolny}
\email{tobias.nadolny@unibas.ch}
\affiliation{Department of Physics, University of Basel, Klingelbergstrasse 82, 4056 Basel, Switzerland}
\author{Christoph Bruder}
\affiliation{Department of Physics, University of Basel, Klingelbergstrasse 82, 4056 Basel, Switzerland}
\author{Matteo Brunelli}
\email{matteo.brunelli@college-de-france.fr}
\altaffiliation[Current address: ]{JEIP, UAR 3573, CNRS, Coll\`ege de France, PSL Research University,
75321 Paris Cedex 5, France\vspace{1em}}
\affiliation{Department of Physics, University of Basel, Klingelbergstrasse 82, 4056 Basel, Switzerland}

\begin{abstract}
Active agents are capable of exerting nonreciprocal forces upon one another.
For instance, one agent, say $A$, may attract another agent $B$ while $B$ repels $A$.
These antagonistic nonreciprocal interactions have been extensively studied in classical systems, revealing a wealth of exciting phenomena such as novel phase transitions and traveling-wave states.
Whether these phenomena can originate in quantum many-body systems is an open issue, and proposals for their realization are lacking.
In this work, we present a model of two species of quantum spins that interact in an antagonistic nonreciprocal way of the attraction-repulsion type.
We propose an implementation based on two atomic ensembles coupled via chiral waveguides featuring both braided and non-braided geometries. 
The spins are active due to the presence of local gain, which allows them to synchronize.
In the thermodynamic limit, we show that nonreciprocal interactions result in a nonreciprocal phase transition to time-crystalline traveling-wave states, associated with spontaneous breaking of parity-time symmetry.
We establish how this symmetry emerges from the microscopic quantum model.
For a finite number of spins, signatures of the time-crystal phase can still be identified by inspecting equal-time or two-time correlation functions.
Remarkably, continuous monitoring of the output field of the waveguides induces a quantum traveling-wave state: a time-crystalline state of a finite-size quantum system, in which parity-time symmetry is spontaneously broken.
Our work lays the foundation to explore nonreciprocal interactions in active quantum matter.
\end{abstract}

\maketitle

\begin{figure*}
    \centering
    \includegraphics[width = 7in]{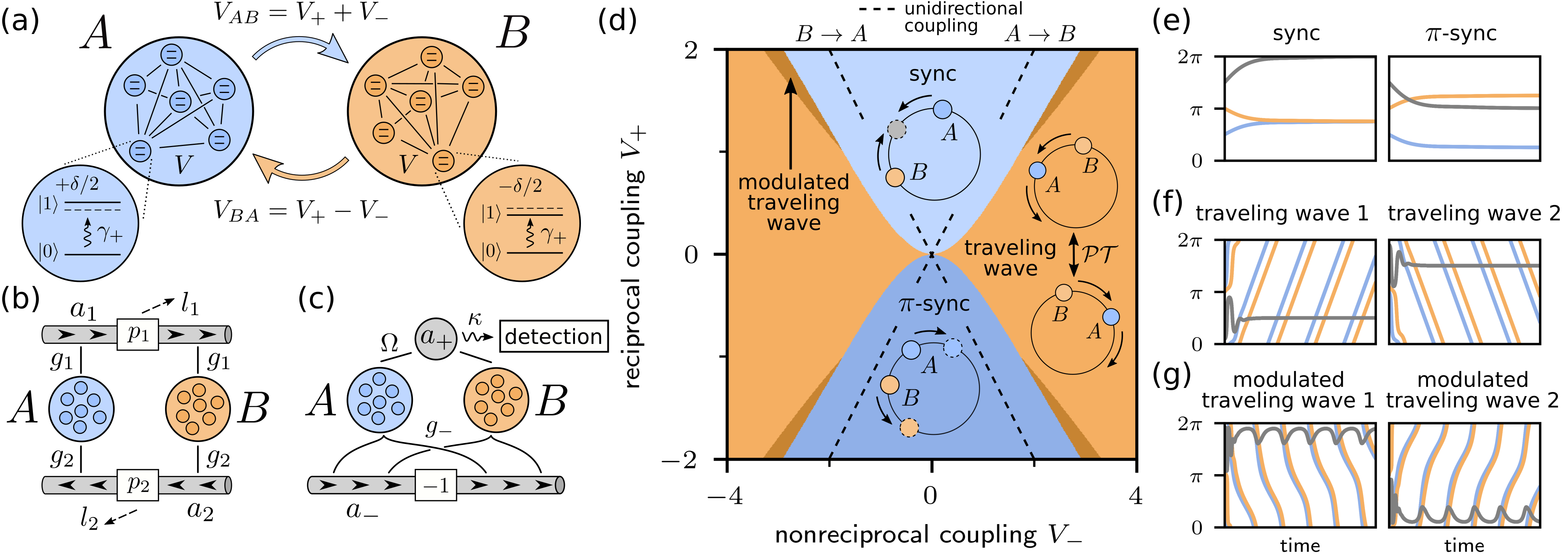}
    \caption{
    (a)
    The model comprises two groups $A$ and $B$ of $N$ quantum spins-$1/2$ each.
    Each spin is incoherently driven from $\ket{0}$ to $\ket{1}$ at rate $\pump$ and the two spin species are detuned by $\delta$.
    Within each species, the spins are coupled with a strength $V$.
    Spins of species $A$ ($B$) influence spins of species $B$ ($A$) with a strength $V_{AB}$ ($V_{BA}$).
    If $V_{AB} \neq V_{BA}$ the interspecies coupling is nonreciprocal.
    (b)
    Physical implementation with doubly-cascaded interactions mediated by two independent chiral waveguides.
    The chiral modes $a_{1,2}$ couple to the spins with strengths $g_{1,2}$.
    Between the two spin species, they pass a phase shifter which multiplies the modes by $p_{1,2} \in \{\pm 1\}$.
    Potential losses $l_{1,2}$ are also accounted for.
    (c)
    Alternative implementation with braided interactions.
    The mode $a_+$ is bidirectional and mediates reciprocal interactions between species $A$ and $B$.
    The chiral mode $a_-$ mediates purely coherent interspecies interactions.
    (d)
    Phase diagram in the thermodynamic limit.
    The insets sketch the dynamics in the respective phase. They depict species $A$ and $B$ as a colored disk whose position on a circle represents the phase of the spins; 
    the arrows indicate the phase velocity.
    Stationary steady-state positions are indicated by disks with a dashed border.
    Parameters: $\delta=0$, $V=2$ (all coupling strengths in units of $\pump$). 
    (e,f,g)
    Time evolution of the phases $\phi_a = \arg[s^+_a]$ of species $A$ (blue) and $B$ (orange) in the thermodynamic limit.
    The phase difference is shown in grey.
    Parameters: $\delta=0$, $V=2\pump$,
    sync: $V_- = 0$, $V_+ = \pump$;
    $\pi$-sync: $V_- = 0$, $V_+ = -\pump$;
    traveling wave: $V_- = \pump$, $V_+ = 0$;
    modulated traveling-wave: $V_- = 2.4\pump$, $V_+ = 1.5\pump$.
    }
    \label{fig:fig1}
\end{figure*}

\section{Introduction}
\label{s:intro}

Active agents are nonequilibrium entities that convert energy into motion or nonconservative forces at the individual level~\cite{Bowick_2022,Shankar_2022}.
A distinctive feature of active agents is that the forces exerted among them can be nonreciprocal. Nonreciprocity between two agents $A$ and $B$ occurs when the response of $A$ to the action of $B$ differs from that of $B$ to $A$.
There are several universal manifestations of nonreciprocal interactions in active matter, \textit{viz}., ensembles of active agents. 
They include the transition from a static to a dynamical behavior~\cite{You_2020},
a new class of critical phenomena marked by exceptional points and spontaneous breaking of parity-time symmetry~\cite{Fruchart_2021},
as well as a dynamical analogue of geometrical frustration resulting in time-crystalline order~\cite{Hanai_2024}.
The effects of nonreciprocal interactions are important in a variety of contexts ranging from pattern formation
\cite{Brauns_2024} to classical spin models~\cite{Avni_2023,Loos_2023}.
They have been observed in metamaterials of various kinds~\cite{Brandenbourger_2019,Librandi_2021,Liu_2024}, nanoparticles~\cite{Reisenbauer_2023}, as well as in active forms of colloids, solids and plasmas~\cite{Meredith_2020,Baconnier_2022,Ivlev_2015}. 
Underlying all of these manifestations is a common type of nonreciprocal interactions, namely \textit{antagonistic interactions} akin to predator-prey dynamics, where $A$ is attracted by $B$ while $B$ is repelled by $A$.
This effect is maximized in the limit of 
interactions with opposite strength but equal magnitude.

Nonreciprocity in quantum systems has also gathered a great deal of interest.
It is a resource for routing information in bosonic networks~\cite{Ranzani_2015, Wanjura_2023} and for quantum sensing~\cite{Lau_2018}.
Quantum emitters coupled to a chiral, i.e., unidirectional, waveguide have been proposed for the preparation of entangled states~\cite{Stannigel_2012, Pichler_2015}. 
In tight-binding models, nonreciprocity is responsible for anomalous localization properties~\cite{HatanoNelson_1996,HatanoNelson_1997,McDonald_2022} and, supplemented with gain, gives rise to exotic non-Hermitian topological phases of bosonic systems~\cite{Wanjura_2020, Porras_2019,Brunelli_2023,Okuma_2023}.
Nonreciprocity is also of high practical relevance. 
Tunable and magnetic-free nonreciprocal quantum devices, implemented with optomechanics~\cite{Verhagen_2017} or superconducting circuits~\cite{Lecocq_2017}, are advantageous for on-chip integration, scaling up superconducting quantum architectures, and improving quantum measurement readout. 
Recently, first steps toward understanding the role of nonreciprocal phase transitions~\cite{Chiacchio_2023,Zhu_2024} and nonreciprocal many-body interactions~\cite{Brighi_2024,Begg_2024} have been taken. 

Current investigations of nonreciprocity in quantum systems are, however, more restrictive than those of nonreciprocal interactions in classical active matter, since only the magnitude, i.e., the absolute value, of the interaction strengths (or transmission coefficients) is typically considered~\cite{Deak_2012, caloz_2018}.
This form of \emph{magnitude nonreciprocity} is maximized for unidirectional interactions, i.e., when $A$ exerts an influence on $B$ while $B$ is immune to $A$’s influence, which corresponds to cascaded quantum dynamics~\cite{Gardiner_1993,Carmichael_1993}.
In contrast, antagonistic interactions imply that $A$'s influence on $B$ and $B$'s influence on $A$ oppose each other, i.e., realize a stronger form of nonreciprocity than unidirectional interactions.
Antagonistic attraction and repulsion is not accounted for by cascaded interactions or other previously studied forms of nonreciprocity in quantum systems, e.g., based on reservoir engineering~\cite{Metelmann_2015}
or dissipative gauge symmetry~\cite{Wang_2023}.
Despite their importance in classical active matter, antagonistic many-body interactions in the quantum regime have remained unexplored.
Furthermore, it is not a priori clear how they can be realized.

In this work, we propose a model of synchronization of active quantum spins to explore the effects of antagonistic nonreciprocity in quantum many-body systems.
Synchronization is the phenomenon of self-organization of coupled oscillators~\cite{pikovsky_rosenblum_kurths_2001}. Nonreciprocal couplings can lead to attraction-repulsion interactions of their phases
which, for classical oscillators, has been shown to result in nonreciprocal phase transitions and traveling-wave states~\cite{Hong_2011_PRL,Sonnenschein_2015,Hanai_2024,Fruchart_2021}.
While synchronization in quantum systems has been the subject of numerous works
\cite{lee_QuantumSynchronizationQuantum_2013,walterQuantumSynchronizationDriven2014,Xu_2013,zhuSynchronizationInteractingQuantum2015,Weiner_2017,Roth_2016,Lorenzo_2022},
the role of antagonistic interactions has not been studied.

The model at the center of this work comprises two species of quantum spins-1/2, as sketched in \cref{fig:fig1}(a).
Each spin is active due to an incoherent gain which provides it with energy.
The spins within each species mutually synchronize due to a collective coupling to a common mode~\cite{zhuSynchronizationInteractingQuantum2015} similar to the phenomenon of superradiance~\cite{Gross_1982}.
The interactions between the two species can be tuned from reciprocal  couplings to unidirectional and antagonistic couplings.
We find that, among active spins, antagonistic interactions are mediated by purely Hamiltonian couplings.
We propose an implementation of this model in a light-matter coupled system consisting of two atomic ensembles coupled by two chiral waveguides which mediate unidirectional couplings each, see \cref{fig:fig1}(b).
Braided interactions reminiscent of the coupling of giant atoms~\cite{Kockum_2018}, as shown in \cref{fig:fig1}(c), can be used to further enhance the antagonistic effects.
In the thermodynamic limit, the model features a nonreciprocal phase transition between static synchronized phases and a dynamical traveling-wave phase where the two species persistently oscillate, see \cref{fig:fig1}(d).
The transition is accompanied by spontaneous breaking of parity-time ($\mathcal{PT}$) symmetry and is induced by antagonistic interactions, thus going beyond the scope of magnitude nonreciprocity.
We show how the nonreciprocal phase transition emerges from an underlying open quantum system description.
Guided by the physical implementation of \cref{fig:fig1}(b), we define a suitable \PTsymm for the Lindblad master equation for any system size that recovers the symmetry that is spontaneously broken in the thermodynamic limit.

We show that, in finite-size quantum systems,
the \PTsymm manifests itself in the spin correlations and the spectral density where an exceptional point signals the nonreciprocal phase transition. 
While in classical systems full knowledge about the state is available, in a quantum system any measurement can only reveal a limited amount of information.
Additionally, the choice of measurement alters the dynamical evolution.
Quantum trajectories reveal that measurement backaction can spontaneously break \PTsymm.
These results show that traveling-wave states and nonreciprocal phase transitions are observable for finite-size quantum systems.

Our work extends the concept of nonreciprocal phase transitions to quantum systems.
From the analysis of the proposed model of active quantum spins that synchronize, we are able to discuss in \cref{s:general_discussion} general implications for a broader class of active quantum systems.
They include the engineering of antagonistic interactions, the importance of \PTsymm on the level of the microscopic quantum master equation, the role of decoherence to stabilize a nonstationary state, and the influence of measurement backaction.
We expect these to be relevant for other models of nonreciprocal phase transitions in active quantum systems.

\section{Quantum model}
\label{s:qumaster}

\subsection{Quantum master equation}
\label{s:sub_qumaster}

The agents of our model are two-level quantum systems, i.e., spins-$1/2$ or simply spins, grouped in two species $A$ and $B$, see \cref{fig:fig1}(a).
For each spin, we define the Pauli $z$-matrix $\sigma^z_{a,i} = \dyad{1}_{a,i} -\dyad{0}_{a,i}$, as well as raising and lowering operators $\sigma^+_{a,i} = \dyad{1}{0}_{a,i}$ and $\sigma^-_{a,i} = \dyad{0}{1}_{a,i}$.
The indices run over the group label $a \in \{A,B\}$ and the $N$ spins per species $i \in \{1,...,N\}$.
All spins within each species are equal and they are described by the collective spin operators $S^{\pm,z}_a = \sum_{i=1}^N \sigma^{\pm,z}_{a,i}$. 

The central quantity of the theory of open quantum systems is the density matrix $\rho$~\cite{Breuer} which in our case describes the state of all spins. 
Its evolution is governed by the quantum master equation
\begin{align}\label{eq:master_spin}
        \dot \rho =
        \mathcal{L} \rho &= 
        -i \left[H_0 + H_\mathrm{inter},\rho \right]
        + (\mathcal{L}_\mathrm{inter}  + \mathcal{L}_\mathrm{intra} + \mathcal{L}_\mathrm{drive})\rho
        \,.
\end{align}
The Liouvillian operator $\mathcal{L}$ is the short-hand notation for the generator of the dynamics.
The dynamics comprises both Hamiltonian terms which generate unitary evolutions
and dissipative terms, which arise from the system being in contact with an (unspecified) environment. The label `inter' refers to \emph{interspecies} coupling, i.e., coupling between spins of different species, while  `intra' refers to \emph{intraspecies} coupling, i.e., coupling within the same species. Here and in the following, we set $\hbar  = 1$.

The Hamiltonian, i.e., coherent, parts are
\begin{equation}
 H_0=
        \delta(S_A^z-S_B^z)/4
        \, , \quad
        H_\mathrm{inter} = 
        i\frac{V_-}{2N} S^+_A S^-_B  + \mathrm{H.c.}
         \raisetag{4\baselineskip}\, .
    \label{eq:master_H_inter}
\end{equation}      
The bare part $H_0$ describes a frequency splitting $\delta$ between the spins of species $A$ and the spins of species $B$.
The interaction part $H_\mathrm{inter}$ describes excitation exchanges with a purely imaginary amplitude, i.e.,  $V_-$ is real valued.
We will later consider the more general case of complex $V_-$ (cf.~\cref{fig:explicit_mf}).
As we will discuss in the following section, $H_\mathrm{inter}$ can be effectively implemented with chiral waveguides and results in antagonistic interactions.
Besides coherent coupling, we include a second type of interspecies coupling of dissipative nature with real-valued strength $V_+$, 
\begin{equation}\label{eq:master_L_inter}
\mathcal{L}_\mathrm{inter}\rho=
\frac{V_+}{N} \left(\mathcal{D}[S^-_A,S^-_B]+\mathcal{D}[S^-_B,S^-_A] \right)\rho \,,
\end{equation}
where $\mathcal{D}[o_1,o_2]\rho = o_1\rho o_2^\dag - (o_2^\dag o_1 \rho + \rho o_2^\dag o_1 )/2$.
Importantly, $V_+$ is allowed to take negative values.
We also include dissipative intraspecies couplings with non-negative strength $V\ge0$,
\begin{equation}\label{eq:master_L_intra}
\mathcal{L}_\mathrm{intra}\rho = \frac{V}{N}
        \left(\mathcal{D}[S^-_A]+\mathcal{D}[S^-_B] \right)\rho\,, 
\end{equation}        
where $\mathcal{D}[o]\rho$ indicates the standard dissipator~\cite{Breuer}, to which $\mathcal{D}[o_1,o_2]\rho$ reduces for equal jump operators $o_1=o_2\equiv o$.
While the terms in Eq.~\eqref{eq:master_L_inter} do not appear to be in standard Lindblad form, the full master equation~\eqref{eq:master_spin} is, as can be seen by combining the two dissipative interactions into collective jump operators using
$\mathcal{D}[o_1\pm o_2] =
\mathcal{D}[o_1] +\mathcal{D}[o_2] 
\pm (\mathcal{D}[o_1,o_2] + \mathcal{D}[o_2,o_1])$.
The combined dissipative interaction terms are
\begin{equation}
\begin{split}
        \mathcal{L}_\mathrm{inter} + \mathcal{L}_\mathrm{intra}  
        =& \frac{\abs{V_{+}}}{N}\mathcal{D} [S^-_A + \mathrm{sign}(V_{+}) S^-_B] +
        \\
        &+ \frac{V - \abs{V_+}}{N} \left( \mathcal{D}[S^-_A]+ \mathcal{D}[S^-_B] \right)
         \, .
    \end{split}
\end{equation}
The master equation describes a 
physical evolution
when all rates of standard dissipators are non-negative.
Therefore, we require $V\geq \abs{V_+}$.

The collective jump operators $S^-_{A,B}$ (and their sum or difference) arise from a collective coupling of all spins (of $A,B$) to a rapidly decaying mode.
This process is known as superradiance~\cite{Gross_1982} and results in a buildup of coherence among the spins.
It can also be understood as synchronization of the spins~\cite{Xu_2013,zhuSynchronizationInteractingQuantum2015}.
Indeed, all three interaction terms contribute to the synchronization dynamics, as we will see in \cref{s:phase_dynamics_and_sync}.

Finally, we consider incoherent driving of each spin at rate $\pump$, described by
\begin{equation}\label{eq:master_L_drive}
\mathcal{L}_\mathrm{drive}\rho= \pump \sum_{i=1}^N \left(\mathcal{D}[{\sigma}_{A,i}^+] +
        \mathcal{D}[{\sigma}_{B,i}^+] \right) \rho\,.
\end{equation}
Unlike the previous terms, the drive acts individually on each spin.
The local gain is crucial as it provides energy at the microscopic level, a distinctive feature of active systems, ranging from lasers to active matter.
In contrast to a thermal bath, the incoherent drive can result in population inversion, where the state $\ket{1}$ is more populated than the state $\ket{0}$, an important feature as we will see in \cref{s:phase_dynamics_and_sync}.

The proposed model possesses different symmetries.
First, the master equation~\eqref{eq:master_spin} is invariant under a global phase shift $\sigma_{a,i} \rightarrow \exp(i\theta)\sigma_{a,i} $ for all spins $\sigma_{a,i}$, ${i\in\{1,...,N\}}$ of both species $a\in\{A,B\}$.
This is a \Usymm, which implies that no phase is preferred.
We will see that phase locking spontaneously breaks the symmetry in the thermodynamic limit in \cref{s:phase_dynamics_and_sync}.
Second, there is no explicit time dependence of the master equation.
The time-translation invariance can also be spontaneously broken in the thermodynamic limit forming a continuous time crystal~\cite{Sacha_2017,Iemini_2018,Kongkhambut_2022}, in analogy to a standard crystal which breaks space-translation invariance, cf.~\cref{s:traveling_wave_states}.
We will point out a third symmetry of \cref{eq:master_spin} in \cref{s:ptsymm}, namely parity-time ($\mathcal{PT}$) symmetry, which implies an invariance of the Liouvillian under complex conjugation $\mathcal{L} = \mathcal{L}^*$ (when $\delta = 0$).

\subsection{Cascaded quantum master equation}
\label{s:cascaded_master_equation}
The model defined by \cref{eq:master_spin} admits a simple interpretation in the framework of cascaded quantum systems~\cite{Gardiner_1993,Carmichael_1993,Stannigel_2012}. 
Cascaded interactions describe unidirectional couplings where one subsystem influences another one but not vice versa. The Liouvillian $\mathcal{L}$ defined in Eq.~\eqref{eq:master_spin} can be rewritten in the equivalent form  
\begin{equation}
    \begin{split}
        \mathcal{L} \rho = &
        - (V_{+}+V_-) ( [S^+_B,  S^-_A\rho] + [\rho S^+_A,S^-_B]) /2N
        \\
      & - (V_{+}-V_-) ( [S^+_A,  S^-_B\rho] + [\rho S^+_B,S^-_A])   /2N
      \\
      & -i[H_0,\rho]
      + \mathcal{L}_\mathrm{intra} \rho
      + \mathcal{L}_\mathrm{drive}\rho
    \, .
    \label{eq:master_casc}
    \end{split}
\end{equation}
The first and second lines each describe cascaded quantum interactions from $A$ to $B$ and $B$ to $A$, respectively, following the notation of~\cite{Stannigel_2012}.
Therefore, we see that
the strength with which species $A$ influences species $B$ is $V_{AB} = V_+ + V_-$ and the strength of $B$'s influence on $A$ is $V_{BA} = V_+ - V_-$, as depicted in \cref{fig:fig1}(a).
When the coupling strengths $V_{AB}$ and $V_{BA}$ differ, the interactions are \textit{nonreciprocal}.
We consequently refer to the parameter $V_- = (V_{AB} - V_{BA})/2$ as nonreciprocal coupling strength 
and to $V_+ = (V_{AB} + V_{BA})/2$ as reciprocal coupling strength.
To summarize, by rewriting the master equation~\eqref{eq:master_spin} in the cascaded form of~\cref{eq:master_casc}, we have found that the nonreciprocal coupling originates from the Hamiltonian term of~\cref{eq:master_H_inter}, while the reciprocal coupling stems from dissipative coupling of~\cref{eq:master_L_inter}.

In the absence of nonreciprocity, $V_- = 0$, the model defined in Eq.~\eqref{eq:master_spin} or \eqref{eq:master_casc} reduces to the model of two atomic ensembles that synchronize via purely dissipative, reciprocal interactions, proposed in Ref.~\cite{Xu_2013}.
Unidirectional synchronization, studied in Ref.~\cite{Roth_2016}, is recovered when either $V_{AB}$ or $V_{BA}$ vanishes.
This configuration maximizes magnitude nonreciprocity, i.e., an asymmetry in the magnitude of the directional couplings, $\abs{V_{AB}} \neq \abs{V_{BA}}$,
which requires both dissipative and coherent interactions~\cite{Metelmann_2015}.
In contrast, our interest is in the even more nonreciprocal scenario in which $V_{AB}$ and $V_{BA}$ take opposite signs, $\mathrm{sign}(V_{AB}) = - \mathrm{sign}(V_{BA})$, which implies that nonreciprocal couplings dominate over reciprocal couplings, $\abs{V_-} > \abs{V_+}$.
In this case, from Eq.~\eqref{eq:master_casc} we see that the maximally nonreciprocal configuration $V_{AB}=-V_{BA}$ is realized by purely coherent interspecies interactions, i.e., $V_+ = 0$.
This may seem surprising at first, 
since a purely Hamiltonian coupling between the two species has to be reciprocal. However, in our model each spin is incoherently driven, see Eq.~\eqref{eq:master_L_drive}.
We shall see that this enables antagonistic quantum interactions of the phases of the spins, thus going beyond magnitude nonreciprocity.
Nonreciprocity in the phase has also been investigated in other contexts. For instance, in linear circuit theory, it is achieved by means of a gyrator~\cite{Tellegen_1948}, i.e., a two-port device that imparts a $\pi$ phase shift only in one direction.
Recently, the use of such nonreciprocal elements has been proposed for the robust encoding of a logical qubit in superconducting circuits~\cite{rymarz_2021}.

\section{Physical implementation}
\label{s:phys_impl_model}

We now suggest a physical realization of the model defined in Eq.~\eqref{eq:master_spin} that can be implemented in current experimental setups.
It is based on the connection with cascaded interactions highlighted in Eq.~\eqref{eq:master_casc} and allows to gain more intuition about our model.
We focus on the interacting part since the detuning and the incoherent drive of each spin are standard and routinely implemented, e.g., in atomic ensembles~\cite{Weiner_2017}.

The simplest configuration that implements our model is depicted in \cref{fig:fig1}(b) and consists of the two spin ensembles $A$ and $B$  coupled by two independent chiral wave\-guides.
Chiral waveguides mediate cascaded, i.e., unidirectional, interactions~\cite{Pichler_2015}, in our case between the two spin species.
Experimentally, such interactions have been implemented between an atomic ensemble and a micromechanical membrane by freely propagating laser beams~\cite{Karg_2020}.
Cascaded interactions between single emitters and receivers have also been engineered in various other platforms~\cite{Petersen_2014,Sollner_2015,Delteil_2017,Joshi_2023}.

The two chiral waveguides in \cref{fig:fig1}(b) allow for two modes $a_1$ and $a_2$ to propagate in opposite directions and interact sequentially with the two spin ensembles $A$ and $B$ with strengths $g_1$ and $g_2$.
Between the two ensembles, the modes pass a phase shifter which transforms $a_{1,2}$ to $p_{1,2} a_{1,2}$.
For simplicity, we focus on a change in sign only, $p_{1,2}\in \{\pm 1\}$.
We also account for the losses $l_{1,2}$ in the chiral waveguides between the two species via the transmission coefficients $0\leq\eta_{1,2} = \sqrt{1-l_{1,2}^2}\leq 1$.

Due to causality,
the mode $a_1$ can only result in species $A$ influencing species $B$, and vice versa for mode $a_2$.
They respectively mediate effective interspecies couplings with strengths 
\begin{equation}
V_{AB}/N=p_1 2g_1^2 {\eta_1} \quad \mathrm{and} \quad V_{BA}/N=p_2 2g_2^2 {\eta_2} \, .
\label{eq:unidirectional_strengths}
\end{equation}
The setup shown in \cref{fig:fig1}(b) is described by the effective master equation \cref{eq:master_spin} or \eqref{eq:master_casc}, 
with
$V_\pm/N = p_1 g_1^2 {\eta_1} \pm p_2 g_2^2 {\eta_2}$
as well as $V/N = g_{1}^2 + g_{2}^2$.
The presence of phase shifts is essential to achieve interspecies couplings with opposite signs, and thus allows for antagonistic interactions.

In this configuration, the interspecies coupling strengths cannot be arbitrarily large compared to the intraspecies coupling strength, since $\abs{V_\pm} \leq V$.
While the constraint $\abs{V_+} \leq V$ is unavoidable for the master equation to be physical, the coherent interactions between the two groups can be enhanced by considering looped or braided geometries known from giant atoms, which allow for $\abs{V_-} > V$~\cite{Kockum_2018,Karg_2019};
a possible implementation is depicted in \cref{fig:fig1}(c).
It makes explicit the distinction between reciprocal and nonreciprocal couplings, mediated by the modes $a_+$ and $a_-$, respectively.
Braided couplings have been demonstrated
in several physical systems, e.g., between atomic spins and a mechanical oscillator~\cite{Karg_2020}, superconducting qubits~\cite{Kannan_2020} and magnetic spin ensembles~\cite{Wang_2022}.

The outputs of the waveguides provide a way to observe the spin dynamics using the input-output relation~\cite{Gardiner_1993}.
The output fields of the two chiral modes shown in \cref{fig:fig1}(b) are
\begin{equation}
\begin{split}
    a_{1,\mathrm{out}} = a_{1,\mathrm{in}} + g_{1}(S^-_A + p_{1} S^-_B) \, ,
    \\
    a_{2,\mathrm{out}} = a_{2,\mathrm{in}} + g_{2}(S^-_B + p_{2} S^-_A) \, .
    \end{split}
    \label{eq:outlightfield}
\end{equation}
They allow to observe correlations among the spins that we will discuss in \cref{s:finitesize}.

\section{Synchronization dynamics}
\label{s:phase_dynamics_and_sync}

To obtain intuition about the dynamics, we first employ a mean-field approach where any correlations between spins are neglected.
This way, each spin (group label $a\in\{A,B\}$ and index $i \in \{1,...,N\}$) has three degrees of freedom.
They are the expectation values of spin operators $\expval{\sigma^{\pm,z}_{a,i}} = \Tr[\sigma^{\pm,z}_{a,i}\rho]$ taken with the density matrix at a given time.
The \textit{coherence} $s^{+}_{a,i} = \expval{\sigma^{+}_{a,i}}$ is the expectation value of the ladder operator $\sigma^{+}_{a,i}$.
The \textit{population} $s^{z}_{a,i} = \expval{\sigma^{z}_{a,i}} = \expval*{\dyad{1}_{a,i} -\dyad{0}_{a,i}}$ quantifies which of the states $\ket{0}$ or $\ket{1}$ is more populated.
From the master equation \eqref{eq:master_spin}, we derive the  time evolution of the coherences and populations for each spin.
Setting ${s}_{a,i}^+ = s_{a,i} \exp(i\phi_{a,i})$, with $s_{a,i} \geq 0$, introduces the real-valued phase of each spin $\phi_{a,i}$, which corresponds to the azimuthal phase on the Bloch sphere.

The time evolution of the phases is given by 
\begin{equation}
\begin{split}
     \frac{\mathrm{d}}{\mathrm{d}t} \phi_{a,i} = \delta_a/2 + \frac{s^z_{a,i}}{2N}
    \sum_{b=A,B}
    \sum_{j=1}^N 
    V_{ba}
    \frac{s_{b,j}}{s_{a,i}} \sin(\phi_{b,j}-\phi_{a,i})
    \, ,
\end{split}
\label{eq:eff_kuramoto}
\end{equation}
where $V_{AA} = V_{BB} = V$, and $\delta_A = - \delta_B = \delta$. These equations establish a connection between the open quantum system in Eq.~\eqref{eq:master_spin} and the nonreciprocal Kuramoto model, which describes the time evolution of all-to-all coupled phase oscillators with nonreciprocal interactions $V_{AB} \neq V_{BA}$~\cite{Fruchart_2021,Hanai_2024}. 

In the Kuramoto model~\cite{Kuramoto1975,Acebron_2005}, the coupling strengths between the phases of oscillators are constant parameters.
A positive coupling $K>0$ between two phase oscillators of the form $\mathrm{d} \phi_1/\mathrm{d}t = K \sin(\phi_2 - \phi_1)$ will lead oscillator 1 towards locking in phase with oscillator 2, i.e., $\phi_1$ is attracted by $\phi_2$.
If $K<0$, however, $\phi_1$ is repelled by $\phi_2$, i.e., oscillator 1 tends to align its phase diametrically opposite to the phase of oscillator 2.  
In contrast, in Eq.~\eqref{eq:eff_kuramoto} the coupling strengths are not constant but depend instead on the instantaneous population $s^z_{a,i}$ and amplitude $s_{a,i}$ of the spins.
The overall sign of the factors multiplying the sine terms determines whether the phases of the spins are attracted or repelled.
The incoherent drive continuously pumps the population of each spin to the state $\ket{1}$, i.e., results in $s^z_{a,i} > 0$.
The dynamics are then determined by the effective interactions of the spins phases.

The mean-field treatment is exact in the thermodynamic limit $N \rightarrow \infty$~\cite{Spohn_1980}.
Exploiting the permutational invariance and setting all spins within each group to be equal $s^{\pm,z}_{a,i}
\equiv s^{\pm,z}_{a}$, we obtain the following dynamical evolution for the mean coherences and populations, 
\begin{subequations}
\begin{align}
  \frac{\mathrm{d}}{\mathrm{d}t} {s}_{A}^+
  =&[(-\pump + i \delta) s^+_{A}
  +V  s^+_{A} s^z_{A} 
  +V_{BA} s^+_B    s^z_{A}] /2
\, , \\
  \frac{\mathrm{d}}{\mathrm{d}t} {s}_{B}^+
  =&[(-\pump - i \delta) s^+_{B}
  +V  s^+_{B} s^z_{B} 
  +V_{AB} s^+_A    s^z_{B}] /2
\, ,
\\
  \frac{\mathrm{d}}{\mathrm{d}t} {s}_{A}^z
  =&
  \pump\left( 1- {s}_{A}^z \right)
  -2V {s}_{A}^+{s}_{A}^- 
  -2V_{BA}\Re[{s}_{A}^+{s}_{B}^-]
  \, ,
 \\
  \frac{\mathrm{d}}{\mathrm{d}t} {s}_{B}^z 
  =&
  \pump\left( 1- {s}_{B}^z \right)
  -2V {s}_{B}^+{s}_{B}^- 
  -2V_{AB}\Re[{s}_{A}^+{s}_{B}^-]
  \, .
\end{align}
\label{eq:MF}
\end{subequations}
We first focus on the effect of intraspecies couplings and the transition to a synchronized state within each species.
Since $V \geq 0$, a positive value of the populations $s^z_{a,i} > 0$
is required for synchronization.
This is achieved by the incoherent drive, which allows for population inversion of the spins, see Eqs.~\eqref{eq:MF}(c,d).
At the same time, the incoherent drive also causes the decay of the coherences, see Eqs.~\eqref{eq:MF}(a,b).
This is the process of decoherence~\cite{Breuer} which diminishes the phase locking and thus competes with the intraspecies interactions.
When the rate of the incoherent drive dominates, the spins therefore converge to an unsynchronized state in which
$s^+_{A,B}=0$ due to the strong decoherence and $s^z_{A,B} = 1$ due to the strong driving.
Above a critical value of the coupling strength $V/\pump$, the unsynchronized solution becomes unstable.
In this regime the spins of each species synchronize as indicated by a finite value of the coherences $s^+_{A,B}$~\cite{Xu_2013, zhuSynchronizationInteractingQuantum2015}. 
This process spontaneously breaks the \Usymm of our model:
both the master equation~\eqref{eq:master_spin} as well as the mean-field equations~\eqref{eq:MF} are invariant under a global phase shift
$s_{A,B}^+ \rightarrow s_{A,B}^+ \exp(i\theta)$ for $\theta \in \mathbb{R}$.
Yet, in the thermodynamic limit the spins synchronize onto a phase that depends on the initial conditions, thus
breaking the \Usymm.
Throughout this work, we consider values of $V/\pump$ large enough such that each species is independently synchronized.
We refer to \cref{app:trivial_to_synch}
for further details on the transition to synchronization.

Once synchronization within each species is achieved, the dynamics of the two species is determined by the interspecies couplings.
The resulting phase diagram obtained by integrating Eqs.~\eqref{eq:MF} is shown in \cref{fig:fig1}(d).
When $V_{AB}$ and $V_{BA}$ are both positive, the phases attract each other and all spins of both species synchronize.
When they are both negative, the phases of any two spins from different species repel each other, such that the two species lock with a phase shift of $\pi$.
This results in two different synchronized regimes, which we call synchronized and $\pi$-synchronized, respectively.
In \cref{fig:fig1}(e) we also show the time evolution of the phases of the two species.

The most remarkable case is that of antagonistic interactions, which occurs when $V_{AB}$ and $V_{BA}$ have opposite signs.
In this case, \cref{eq:eff_kuramoto} predicts that the spins of one species, say $A$, try to lock their phases with those of species $B$.
The spins of species $B$, however, have the opposite inclination, namely they tend to lock with a phase difference of $\pi$ with respect to species $A$.

We stress that the incoherent drive plays a key role in enabling persistent nonreciprocal interactions among the phases of the spins, i.e., interactions whose effect extend beyond the transient.
To illustrate this, we adopt a `general notion of nonreciprocity'~\cite{Bowick_2022,Fruchart_2021}, since the dynamics are not described in terms of forces.
We say that two dynamical variables $x_A, x_B$ interact in a nonreciprocal way whenever the coupling coefficients are asymmetric:
$\dot x_A = C_{BA} x_B$, $\dot x_B = C_{AB} x_A$, where $C_{AB} \neq C_{BA}$.
In our model, this scenario
is realized
at the level of the phases (see~\cref{eq:eff_kuramoto} and the discussion below it), or equivalently
at the level of the coherences, as we will now show.
We recast \cref{eq:MF} in matrix form (here setting $\delta = 0$ for simplicity)
\begin{equation}
\frac{\mathrm{d}}{\mathrm{d}t}
    \begin{pmatrix}
        s^+_A\\s^+_B
    \end{pmatrix}
    =
    \frac{1}{2}
    \begin{pmatrix}
        -\pump + V s^z_A  &  V_{BA}  s^z_A  \\
         V_{AB} s^z_B   & -\pump  + V s^z_B 
    \end{pmatrix}
    \begin{pmatrix}
        s^+_A\\s^+_B
    \end{pmatrix}
    \, .
    \label{eq:coherence_matrix}
\end{equation}
After some transient evolution, both $s^z_A$ and $s^z_B$ settle to a positive value due to the incoherent drive.
Therefore, when $V/\gamma_+$ is large enough, the coherences take a finite value due to the synchronization within each species (diagonal part of \cref{eq:coherence_matrix}). 
From the off-diagonal part of \cref{eq:coherence_matrix}, we see that the sign of $V_{AB}$ and $V_{BA}$ determines the quality of the interactions.
The interactions are nonreciprocal when $V_{AB} s^z_A \neq V_{BA} s^z_B$.
Since the species are otherwise identical, this scenario occurs when $V_{AB} \neq V_{BA}$. 
Specifically, if $\sign V_{AB} = - \sign V_{BA}$, they are of the antagonistic type.
A similar argument can be made for any pair of two spins, one of species $A$ and one of species $B$, concluding that the microscopic interactions are nonreciprocal.
In the absence of the incoherent drive, the dynamics of the phases may still be nonreciprocal at special times; however, the incoherent drive allows to sustain nonreciprocal interactions (and the resulting traveling-wave states) in the long-time limit.

\section{Traveling-wave states}
\label{s:traveling_wave_states}

For large enough nonreciprocity, we find a regime of nonstationary states, see \cref{fig:fig1}(d).
They are characterized by stable oscillations of $s^+_a$ in the long-time limit, i.e., the phases $\arg(s^+_a)$ grow linearly in time, see \cref{fig:fig1}(f).
The spins of one species persistently chase after the spins of the other species, which in turn run away.
The phase difference between the two species as well as $\abs{s^+_a}$ and $s^z_a$ are constant.
Such states are called \textit{traveling-wave} states~\cite{Hong_2011_PRL,Fruchart_2021,Brauns_2024}.
From \cref{fig:fig1}(d) we also see that unidirectional interactions $V_+=\pm V_-$ highlighted by the diagonal dashed lines lie entirely inside the static regions which confirms that traveling-wave states are beyond the reach of magnitude nonreciprocity.
Moreover, between the ($\pi$)-synchronized and the traveling-wave regimes, we find a region of modulated traveling-wave states~\cite{Fruchart_2021}, see \cref{fig:fig1}(d).
In these states, the relative phase difference as well as $\abs{s^+_a}$ and $s^z_a$ also oscillate in time.
Exemplary trajectories are shown in \cref{fig:fig1}(g).

The traveling-wave state displays time-crystalline order~\cite{Hanai_2024} and can be understood as an instance of a continuous time crystal~\cite{Sacha_2017,Iemini_2018,Kongkhambut_2022}.
As a time-dependent oscillatory state,
it breaks the time-translation symmetry of the microscopic model~\eqref{eq:master_spin}, analogously to a standard crystal which breaks space-translation symmetry.
The transition from a synchronized state to a traveling-wave state is a nonreciprocal phase transition.
It is known to occur for ensembles of classical phase oscillators, where the phase diagram is similar to that of \cref{fig:fig1}(d)~\cite{Fruchart_2021}.
The emergence of a nonreciprocal phase transition and traveling-wave states from a microscopic quantum spin model is a key finding of our work.

\subsection{Order from decoherence}

For classical phase oscillators, it has been shown that traveling-wave states require the presence of disorder in their frequencies or additional phase noise~\cite{Fruchart_2021, Hanai_2024}.
In their absence, 
the oscillators will eventually reach either a synchronized or a $\pi$-synchronized state.
Adding noise or disorder prevents relaxation towards these stationary states, leading to persistent dynamical states.

In our model, quantum decoherence is responsible for the stabilization of the traveling-wave state, without the need to include any external source of noise or disorder.
Decoherence implies a decay of the coherences $s^+_{A,B}$, see Eqs.~\eqref{eq:MF}(a,b).
While the decoherence resulting from the dissipative interactions defined in Eqs.~\eqref{eq:master_L_inter},~\eqref{eq:master_L_intra} is negligible for a large number $N$ of spins, the incoherent drive at rate $\pump$ entails decoherence independently of $N$.
The decoherence prevents a fully coherent, stationary, state and thus takes the role of noise or disorder to perturb stationary states in classical systems.
We conclude that the incoherent drive is essential for the emergence of traveling-wave states in two ways: it activates each spin by inverting the populations,
which allows for nonreciprocal phase interactions; and it provides decoherence that stabilizes the dynamical state.

\subsection{Spontaneous \texorpdfstring{\PTsymm}{PT symmetry} breaking}
\label{traveling_wave_states_spont}
In the traveling-wave phase, we find two stable solutions that are shown in \cref{fig:fig1}(f).
They differ by the chirality of the emergent oscillation.
Depending on the chirality, the observed frequency is either positive or negative and the phase difference between the two species takes a constant value of approximately $+ \pi/2$ or $- \pi/2$.
The two solutions are related by a symmetry transformation of the mean-field equations~\eqref{eq:MF}.
This symmetry is referred to as a generalized $\mathcal{PT}$ 
(parity-time) 
symmetry.
Introducing a shorthand notation for the mean-field equations~\eqref{eq:MF}, $\mathrm{d} \mathbf{s} /\mathrm{d}t = \mathbf{L}(\mathbf{s})$, where $\mathbf{s} = (s^+_A, s^+_B, s^z_A, s^z_B)$ and $\mathbf{L}(\mathbf{s})$ is the right-hand side of equations~\eqref{eq:MF},
we have
\begin{equation}
    \mathbf{L}(\mathbf{s}) = \mathbf{L}^*(\mathbf{s})\, , \quad \mathrm{if} \, \delta = 0\, .
    \label{eq:pt_symmetry_mean_field_1}
\end{equation}
The \PTsymm implies that when $\mathbf{s}$ is a solution, then $\mathbf{s}^*$ is a solution as well:
\begin{equation}
    \text{for any solution }\, \mathbf{s} \Rightarrow 
    \mathbf{s}^* \text{ is solution}
    \, .
    \label{eq:mirror_symmetry}
\end{equation}
Each of the two stable solutions spontaneously breaks \PTsymm, and the two solutions are related by the symmetry transformation, i.e., complex conjugation.
The spontaneous breaking of \PTsymm is a key feature of nonreciprocal phase transitions~\cite{Fruchart_2021}.
Time-translation invariance is concurrently broken: 
each of the traveling-wave states forms one of two possible time-crystals. 

\subsection{Explicit \texorpdfstring{$\mathcal{PT}$-symmetry}{PT-symmetry} breaking}
\label{s:explicit_breaking_delta}
\begin{figure}
    \centering
    \includegraphics[width = 3.4in]{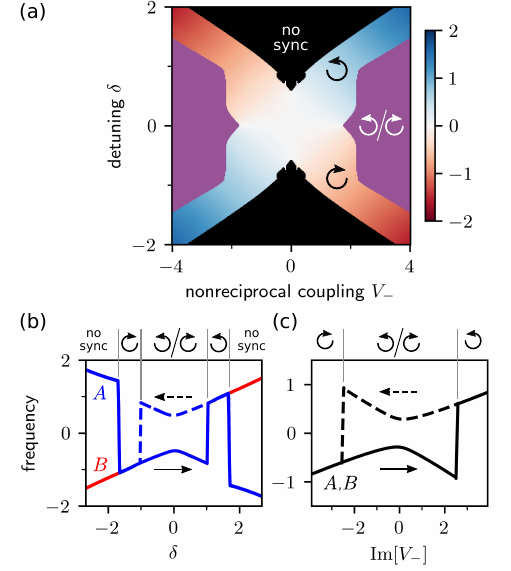}
    \caption{
    Explicit symmetry breaking.
    (a)
    The color bar indicates the frequency of the oscillation of the spins in the case of explicit symmetry breaking ($\circlearrowleft$ or $\circlearrowright$).
    Furthermore, the purple color indicates where the symmetry is spontaneously broken and either of two frequencies is obtained ($\circlearrowleft\hspace{-2pt}/\hspace{-2pt}\circlearrowright$).
    Black shows the unsynchronized regime where the two spin species oscillate at different frequencies.
    (b)
    Cut through (a) at $V_- = 2.5$.
    Shown are the frequencies of species $A$ (red) and $B$ (blue) when increasing (solid line) and decreasing (dashed line) the detuning.
    If the two species are synchronized, their common frequency is shown in blue.
    There are five regions from left to right: an unsynchronized state, a synchronized state in which the symmetry is explicitly broken ($\circlearrowright$), two states that spontaneously break the symmetry (${\circlearrowleft\hspace{-2pt}/\hspace{-2pt}\circlearrowright}$), explicitly broken ($\circlearrowleft$) and unsynchronized.
    (c) Similar to (b) when the symmetry is explicitly broken by $\Im[V_-]$ at constant $\mathrm{Re}[V_-]=1.5$.
    There are no unsynchronized states in this case.
    Parameters: $V_+ = 1$, $V=2$.
    All frequencies and interaction strengths are given in units of $\pump$.
    }
\label{fig:explicit_mf}
\end{figure}

We illustrate the effect of explicitly breaking \PTsymm in \cref{fig:explicit_mf}(a) by considering a finite detuning $\delta$ between the species which was set to zero in the analysis so far.
For small values of $\delta$, spontaneous breaking still occurs as indicated by the purple regime.
This shows that the traveling-wave phase possesses some degree of robustness to frequency imbalance, i.e., the spontaneous breaking of \PTsymm occurs even when the \PTtrans is not a perfect symmetry of the model.
For larger values of $\delta$, however, one of the two traveling-wave states becomes the only stable solution indicated by blue and red regions.
For very large detuning, we find yet another regime (black) in which the two species are unsynchronized and oscillate at different frequencies.

In the traveling-wave regime, oscillations with opposite chiralities are stable.
The bistability is associated with a hysteresis in the observed frequency of the oscillations.
In \cref{fig:explicit_mf}(b), we show the hysteretic behavior by adiabatically increasing and decreasing the detuning.
Starting from a large negative value of $\delta$ where the spins are unsynchronized, 
the spins first enter a traveling-wave phase with unique chirality.
Upon further increasing the detuning, the state maintains the same chirality throughout the coexistence phase, until it becomes unstable and the traveling-wave state with opposite chirality is attained.
When reversing this parameter sweep, however, the spins remain in the state with opposite chirality in the coexistence regime. 
Similarly, including a nonzero imaginary part of $V_-$ also breaks \PTsymm explicitly; in Eq.~\eqref{eq:master_H_inter} the amplitude was chosen to be purely imaginary, i.e., $\Im[V_-]=0$.
We find a hysteretic behavior, as shown in \cref{fig:explicit_mf}(c).
Notice that the unsynchronized state only occurs for large $\delta$, while for large $\Im[V_-]$, the system remains in the explicitly broken regime.

\section{Origin of \texorpdfstring{\PTsymm}{PT symmetry}}
\label{s:ptsymm}
So far we have discussed the generalized \PTsymm as a property of the mean-field equations.
We now show how the notion of \PTsymm is grounded in physical terms and how it emerges from a microscopic theory.
While typically \PTsymm is discussed on the level of non-Hermitian Hamiltonians~\cite{Bender_2023}, here, an understanding of \PTsymm on the level of the master equation is required.
In open quantum systems a distinction can be made between weak and strong symmetries~\cite{Buca2012_symmetry, Albert2014}. 
\PTsymm of master equations has been defined both in a weak~\cite{Prosen_2012,Huybrechts_2020,Sa2023} and in a strong sense~\cite{Huber_2020,Nakanishi_2022}. The notion we put forward corresponds to a weak symmetry.
We will say that a Liouvillian $\mathcal{L}$ is $\mathcal{PT}$-symmetric or invariant under $\mathcal{PT}$
transformation when $\mathcal{PT}\, \mathcal{L} (\mathcal{PT})^{-1}=\mathcal{L}$.
This choice of \PTsymm is physically motivated by the proposed  implementation of our model, as we now show.
\begin{figure}
    \centering
    \includegraphics[width = 3.4in]{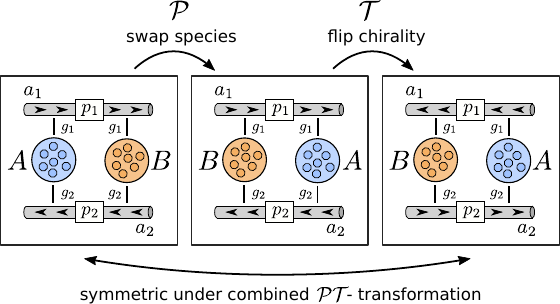}
    \caption{
    Effect of parity ($\mathcal{P}$) and time reversal ($\mathcal{T}$) transformations in the implementation of our model using chiral waveguides.
    The model is $\mathcal{PT}$ symmetric if $\delta = 0$.
    A finite detuning distinguishes the left and right locations of the spins and thereby explicitly breaks the \PTsymm.
    \label{fig:ptsymm}
    }
\end{figure}

The effects of $\mathcal{PT}$  transformation in our model are depicted in \cref{fig:ptsymm}.
In the context of two interacting species, it is typical to define the \textit{parity} transformation ($\mathcal{P}$) as the exchange of the two species $\mathcal{P}: A \leftrightarrow B$~\cite{Huber_2020,Nakanishi_2022,Bender_2023}.
On the master equation level, this corresponds to the unitary swap $\sigma^{\pm,z}_{A,i}  \leftrightarrow \sigma^{\pm,z}_{B,i}$.
If a detuning is present, $A$ and $B$ change their frequency under $\mathcal{P}$.
Therefore, in the implementation shown in \cref{fig:ptsymm}, the detuning is a local property that distinguishes the left and right sides.
For the definition of \textit{time reversal} ($\mathcal{T}$), we inspect the effect of
a time-reversed propagation of the two modes $a_{1,2}$, which is equivalent to flipping the chirality of the waveguides.
Since $V_-/N = p_1 g_1^2 {\eta_1} - p_2 g_2^2 {\eta_2}$, 
this effectively changes the sign of the coherent interactions between the two spin species.
We therefore find
$\mathcal{T}: H_\mathrm{inter} \rightarrow - H_\mathrm{inter}$.
Combining the two transformations, we obtain that, if $\delta = 0$, our model \eqref{eq:master_spin} is $\mathcal{PT}$ symmetric, i.e., invariant under the combined \PTtrans ($A \leftrightarrow B$ and $H_\mathrm{inter} \rightarrow - H_\mathrm{inter}$).
Otherwise, \PTsymm is explicitly broken.
An imaginary part of $V_-$ similarly explicitly breaks \PTsymm.
The same transformation is obtained when considering the implementation shown in \cref{fig:fig1}(c), where the bidirectional waveguide is invariant under $\mathcal{T}$, see \cref{app:general_Vm}.

Remarkably, the \PTtrans just introduced is equivalent to complex conjugation of the Liouvillian superoperator $\mathcal{L}$ in Eq.~\eqref{eq:master_spin}, namely 
\begin{equation}
    \mathcal{PT} :
    (A \leftrightarrow B \text{ and }
    H_\mathrm{inter} \rightarrow - H_\mathrm{inter})
    \iff 
    \mathcal{L} \rightarrow \mathcal{L}^* 
    \, .
    \label{eq:pt}
\end{equation}
Complex conjugation of a generic Liouvillian, given a Hamiltonian $H$ and a set of jump operators $\{L_{\mu}\}$, is defined as
$\mathcal{L}\rho=
-i[H,\rho] + \sum\nolimits_{\mu} \mathcal{D}[L_{\mu}  ] \rho
\rightarrow
\mathcal{L}^*\rho=+i[H^*,\rho] + \sum\nolimits_{\mu} \mathcal{D}[L_{\mu}^*] \rho
$.
A $\mathcal{PT}$-symmetric Liouvillian has several important consequences.
First of all, in the thermodynamic limit, the \PTsymm~\cref{eq:mirror_symmetry} of the mean-field equations~\eqref{eq:MF} follows
\footnote{
For any solution $\rho_0$, i.e., $\dot\rho_0 = \mathcal{L} \rho_0$, $\rho_0^*$ is a solution as well, since $\dot \rho_0^* = \mathcal{L}^*\rho_0^*=\mathcal{L}\rho_0^*$.
The mean-field Ansatz used to obtain \eqref{eq:MF} is
\begin{equation*}
    \rho_0(\mathbf{s}) = 
    \bigotimes_{i=1}^N 
    \bigotimes_{a=A,B}
    \left(\mathbb{I}+ s^z_{a} \sigma^z +
    2s^+_{a} \sigma^- +
    2s^-_{a} \sigma^+
    \right)/2
    \, ,
\end{equation*}
where $\mathbb{I} = \dyad{0}+\dyad{1}$.
Given $\mathbf{s}_0$ such that $\rho_0(\mathbf{s}_0)$ is a solution, 
$\rho_0^*(\mathbf{s}_0)=\rho_0(\mathbf{s}^*_0)$ is a solution as well, which implies that both $\mathbf{s}_0$ and $\mathbf{s}^*_0$ solve the mean-field equations~\eqref{eq:MF}.
We have thus recovered the \PTsymm~\cref{eq:mirror_symmetry} in the thermodynamic limit.
}:
\begin{equation}
    \mathcal{L}  = \mathcal{L}^*
    \Rightarrow
    \left(
    \text{for any solution }\, \mathbf{s} \Rightarrow 
    \mathbf{s}^* \text{ is solution}
    \right)
    \label{eq:pt_symmetry_mean_field}
\end{equation}
We have thus established the emergence of \PTsymm, which is spontaneously broken in the thermodynamic limit, from a physical and microscopic theory.
Secondly, as we show in the next section, the symmetry constrains the spin correlations and dynamics for finite-size systems away from the thermodynamic limit.

\section{Finite-size system}
\label{s:finitesize}

\subsection{Steady-state correlations}
\label{s:finitesize_steady_state_equal_time}

We start the analysis of finite-size systems by considering the long-time limit of Eq.~\eqref{eq:master_spin}.
For any finite number of spins, a unique time-independent steady-state density matrix is obtained.
This is a general property of finite-size systems and contrasts the previous analysis in the thermodynamic limit, where a time-dependent state is obtained.
To analyze the steady state, we compute expectation values of operators $\expval{o} = \Tr[\rho o]$.
Since all spins within each species are identical, we can drop the index labeling the individual spins, e.g.,
$\expval{\sigma^z_{A}}=\expval{\sigma^z_{A,i}}$ or 
$\expval{\sigma^+_{A} \sigma^-_{A}}=
 \big\langle \sigma^+_{A,i} \sigma^-_{A,j}\big \rangle$ for $i\neq j$.
The correlations within each species and between species are measurable by considering the output of the chiral waveguides, see \cref{eq:outlightfield}.

The steady state respects the symmetries of the master equation, i.e., \Usymm and \PTsymm.
As a result of \Usymm, expectation values that are affected by a global phase shift, such as $\expval{\sigma^+_{A}}$, vanish, while \PTsymm of Eq.~\eqref{eq:pt} enforces that the unique steady state is real valued, namely
\begin{equation}
    \mathcal{L} = \mathcal{L}^* 
    \Rightarrow 
    \rho_\mathrm{ss} = \rho_\mathrm{ss}^*
    \, .
    \label{eq:ss_density_real}
\end{equation}
Therefore, the expectation value of certain operators, such as $\sigma^+_A\sigma_B^-$, must also be real valued.
We will find this to be important later on.

We compute the exact steady state of Eq.~\eqref{eq:master_spin} for small numbers of spins up to $N=19$ (and $N=90$ for $V_-=0$ and $V_+ = V$, where all spins of both species are equal).
For larger $N$, we resort to the cumulant expansion approximation, which allows to systematically include correlations up to a certain order~\cite{cumulants_kubo}.
Here, we perform an expansion to second and to fourth order, see \cref{app:finite_size_calculations} for details.

\begin{figure}
    \includegraphics[width = 3.4in]{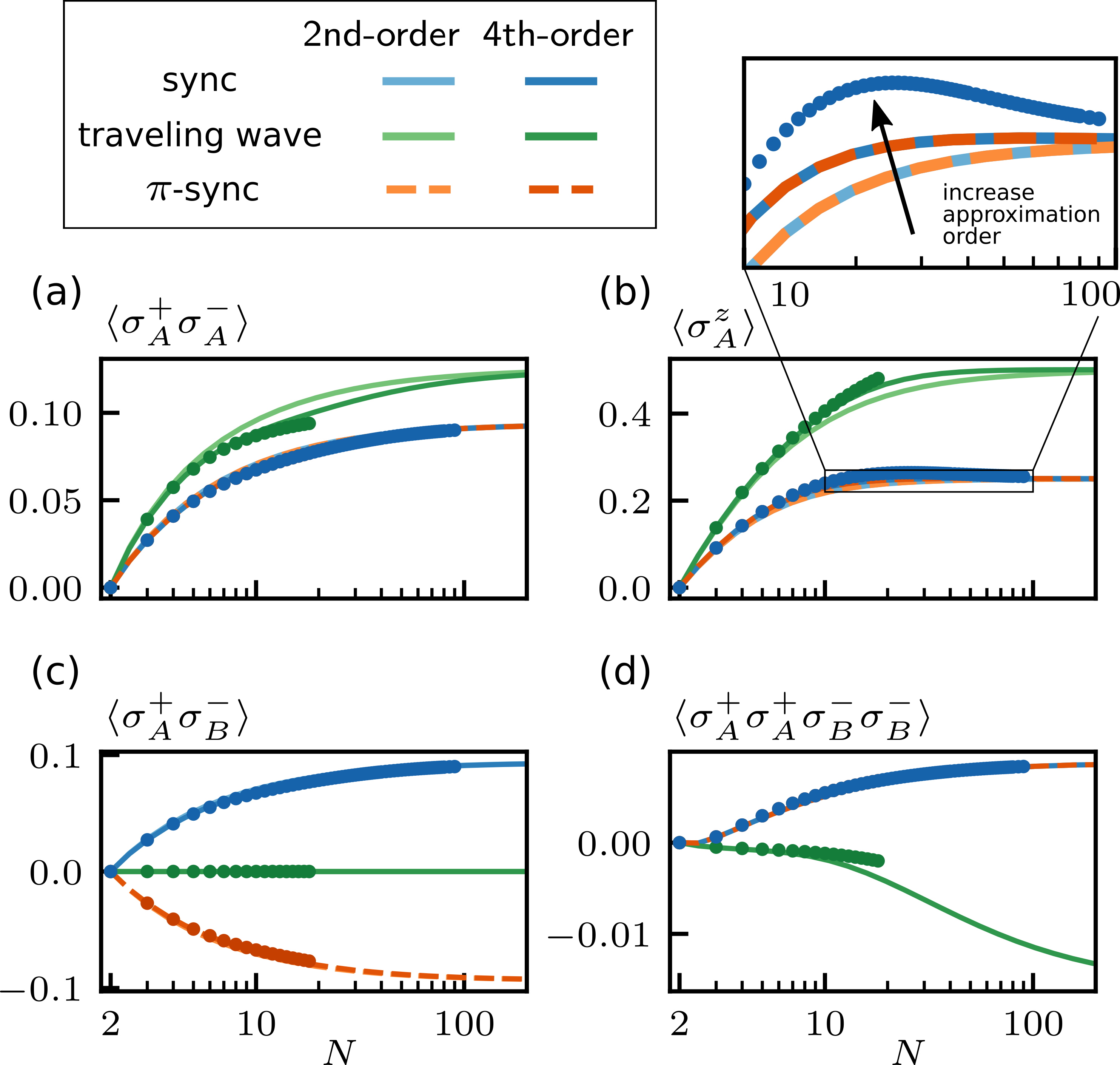}
    \caption{
    Spin correlations and population as a function of number of spins $N$ for $(\pi)$-synchronized and traveling-wave state.
    The solid and dashed lines are computed with the cumulant expansion and the dots (same colors) show the exact solution of the master equation.
    The inset shows a zoom of the nonmonotonous behavior of the populations and highlights the convergence of the approximations.
    Parameters: $\delta=0$,
    $V = 2\pump$,
    synchronized state: $V_-=0, V_+=2\pump$,
    traveling-wave state: $V_-=2\pump, V_+=0$,
    $\pi$-synchronized state: $V_-=0, V_+= -2\pump$. 
    }
    \label{fig:plot_over_N}
\end{figure}

In \cref{fig:plot_over_N}, we show correlations and the populations as a function of the number of spins per species $N$.
The results from the cumulant expansion agree very well with the exact solution for large and small $N$.
For large $N$, the agreement is expected as the cumulant expansion is exact in the thermodynamic limit.
For small $N$, the agreement results from the simplicity of the state that can be well captured by the approximation.
In an intermediate regime, $5 \lesssim N \lesssim 100$, however, there are deviations.
They hint at the existence of nontrivial higher-order correlations in this regime which are not well captured by a low-order cumulant expansion.
We expect the cumulant expansion to converge to the exact results when including higher-order correlations.
Indeed, the fourth-order expansion agrees better with the exact solution than the second-order expansion, as highlighted in the inset of \cref{fig:plot_over_N}(b).

\begin{figure}
    \centering
    \includegraphics[width = 3.4in]{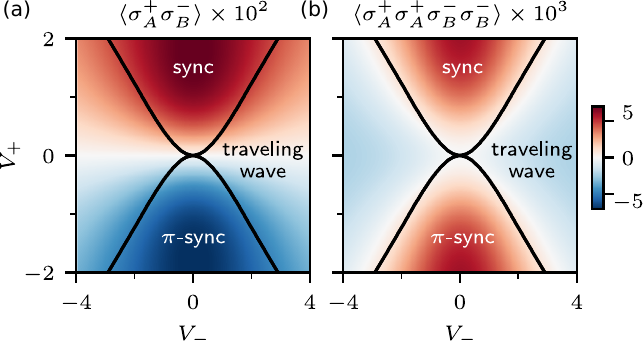}
    \caption{
    (a)
    Second-order correlations $\expval{\sigma^+_A \sigma^-_B}$ between the two groups.
    This quantity allows to distinguish between the synchronized and $\pi$-synchronized phases.
    The correlations are suppressed in the traveling-wave phase.
    The black lines are the boundaries of the mean-field phase diagram in \cref{fig:fig1}(d).
    (b)
    Fourth-order correlations $\expval{\sigma^+_A \sigma^+_A \sigma^-_B \sigma^-_B}$.
    A negative value indicates the traveling-wave state.
    Parameters: $V=2$, $\delta = 0$, all interaction strengths in units of $\pump$ and $N=12$.
    }
    \label{fig:correlations}
\end{figure}

The intraspecies correlations, as measured by $\expval{\sigma^+_{A}\sigma^-_{A}}$, indicate synchronization within each ensemble (an analogous plot is obtained for species $B$).
The correlations increase with the number of spins but do not change significantly beyond $N \gtrsim 200$.
The interspecies correlations $\expval{\sigma^+_A \sigma^-_B}$ quantify the synchronization between species.
In general, $\expval{\sigma^+_A \sigma^-_B}$ is a complex number whose argument determines the relative phase difference between the two species.
In our case, however, \PTsymm implies that these correlations are real valued.
\Cref{fig:plot_over_N}(c) highlights that in the synchronized state the phase difference is zero, while the $\pi$-synchronized state is indicated by negative correlations.
This is clear from \cref{fig:correlations}(a), which shows the correlations as a function of $V_\pm$ at fixed $N$.

In the traveling-wave state, the correlations between the two species vanish when $V_+ = 0$, see \cref{fig:plot_over_N}(c) and \cref{fig:correlations}(a).
However, in the thermodynamic limit
$\expval{\sigma^+_A \sigma^-_B} \xrightarrow{N\rightarrow\infty} s^+_A s^-_B$, we expect one of two possible nonzero complex values, each corresponding to one of two possible solutions related by \PTtrans, i.e., complex conjugation.
The apparent discrepancy is resolved by considering that the density operator describes an ensemble of possible trajectories which contains both spontaneously broken states.
In the two traveling-wave states, the species lock at a phase difference close to $+\pi/2$ or $-\pi/2$.
The average of $s^+_A s^-_B \sim \exp(\pm i \pi/2)$ is therefore small and vanishes for $V_+ = 0$.
We will explicitly show the spontaneous breaking of \PTsymm in quantum trajectories in \cref{s:qu_trajectories}.

Whereas features of the traveling-wave states are not revealed by second-order correlations, they become apparent in higher-order correlations.
We show in \cref{fig:plot_over_N}(d) and \cref{fig:correlations}(b) the fourth-order correlations
$\expval{\sigma^+_{A}\sigma^+_{A}\sigma^-_{B}\sigma^-_{B}}$ between four different spins.
It is negative only in the traveling-wave regime.
In the two possible traveling-wave states of the thermodynamic limit, the species maintain a phase difference close to $\pm \pi/2$.
In both cases the product
$s^+_{A}s^+_{A}s^-_{B}s^-_{B} < 0$
is negative.
Therefore, averaging over the two possible traveling-wave states does not result in a suppression of these correlations.
For any finite-size system, we have thus established the fourth-order correlations as a measure that signals the traveling-wave phase even from the stationary steady-state density matrix.
However, this does not prove the presence of dynamics in finite-size systems yet, and this point will be addressed in the next section.

\subsection{Two-time correlations, spectra, and exceptional points}
\label{s:finitesize_correlations_spectra}

The dynamics of the spins in the finite-size system can be observed through the two-time correlations 
$\langle{\sigma}_{a}^+({t+\tau}){\sigma}_{b}^-(t)\rangle$.
They quantify correlations between spins of species $a,b \in \{A,B\}$ at two times that differ by $\tau$.
To analyze the dynamics in frequency space, we introduce the spectral density for each species defined as the Fourier transform of two-time correlations
\begin{equation}
    P_{a}(\omega) =
    \lim_{t\rightarrow \infty}\int_{0}^\infty \mathrm{d}\tau \,
    \langle{\sigma}_{a}^+(t+\tau){\sigma}_{a}^-(t)\rangle e^{i\omega \tau}
    \, .
    \label{eq:spectrum}
\end{equation}
The spectrum is experimentally accessible via standard methods like heterodyne measurements~\cite{Wiseman_2009} of the output of the chiral waveguides shown in~\cref{fig:fig1}(b) and defined in \cref{eq:outlightfield}.

To calculate the spectrum efficiently, we employ the quantum regression theorem~\cite{Breuer}.
We define a vector $\mathbf{c}$ whose first and second entries are the two-time correlations between any two different spins within species~$A$, $\langle{\sigma}_{A}^+(t+\tau){\sigma}_{A}^-(t)\rangle$, and the ones between two spins of different groups, $\langle{\sigma}_{B}^+(t+\tau){\sigma}_{A}^-(t)\rangle$.
The correlations evolve following
$
\dv*{\tau} \mathbf{c} = M\mathbf{c}
$
with
\begin{equation}
M= \frac{1}{2}
    \begin{pmatrix}
        -\gamma_A +i\delta + V s^z_A  &  V_{BA}  s^z_A  \\
        V_{AB} s^z_B   & -\gamma_B -i\delta  + V s^z_B 
    \end{pmatrix}
    \, ,
    \label{eq:Mmatrix}
\end{equation}
where the effective decoherence rate $\gamma_{A,B} = \pump + V{(1+s^z_{A,B})}/N$ depends on the number of spins $N$.
Further details on the derivation can be found in \cref{app:twotime}.
To compute the spectrum,
we solve Eqs.~\eqref{eq:spectrum} and \eqref{eq:Mmatrix} and set the initial condition $\mathbf{c}(0) = (\langle{\sigma}_{A}^+(t){\sigma}_{A}^-(t)\rangle, \langle{\sigma}_{B}^+(t){\sigma}_{A}^-(t)\rangle)$ to the steady-state correlations obtained from the cumulant expansion discussed in \cref{s:finitesize_steady_state_equal_time}.

\begin{figure}
    \centering
    \includegraphics[width = 3.1in]{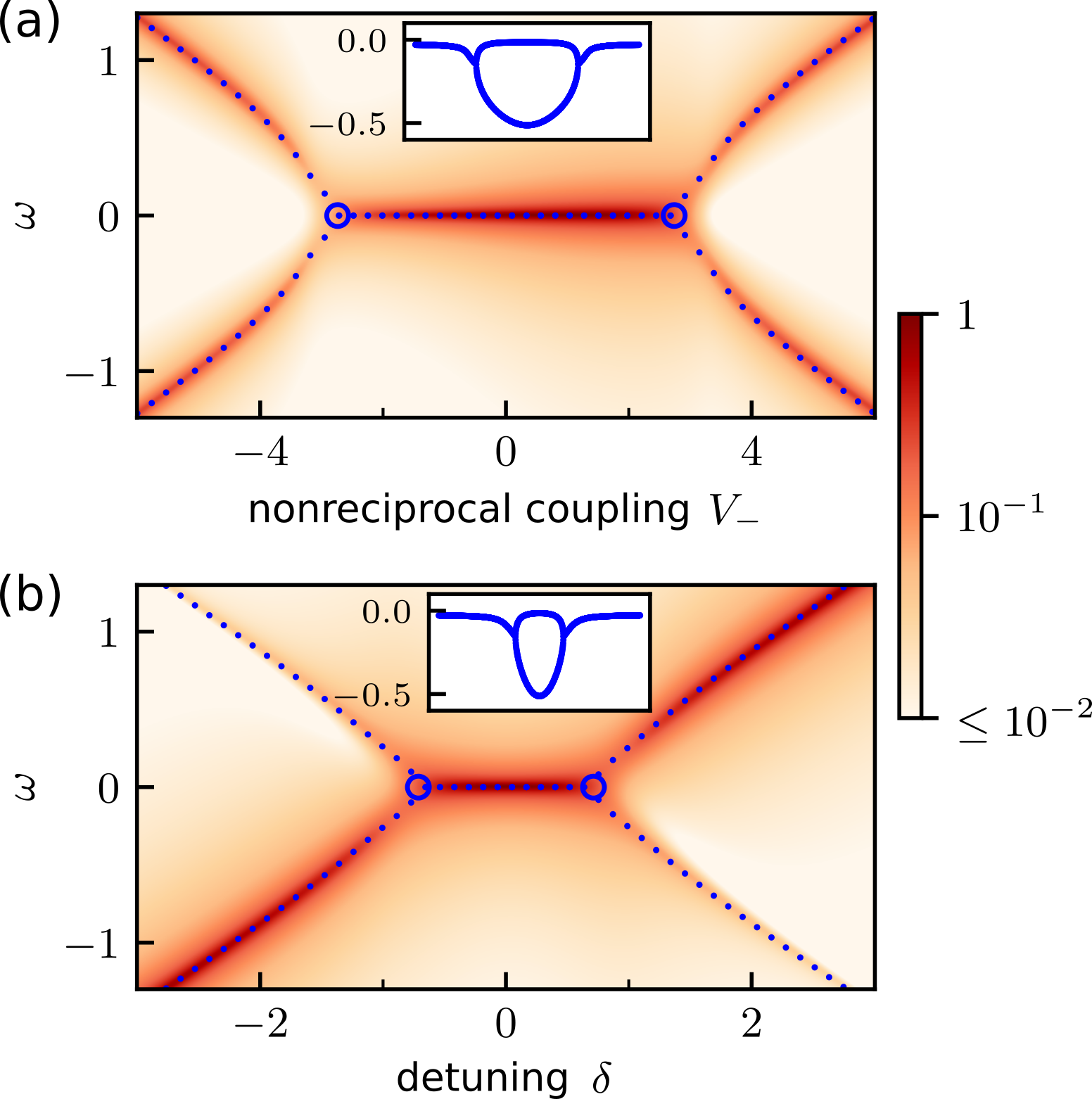}
    \caption{Spectrum $\abs{P_{A}(\omega)}$
    for the finite-size system with $N=100$
    as a function of $V_-$ (a) and $\delta$ (b).
    The spectrum is obtained by using \cref{eq:Mmatrix} together with the cumulant expansion.
    It is shown on a logarithmic scale normalized to the maximum value.
    The blue dots mark the imaginary parts of the eigenvalues of the dynamical matrix $M$.
    Blue circles denote exceptional points.
    The insets show the real part of the eigenvalues of $M$.
    Parameters: $N=100$, $V = 2$, (a) $\delta=0$, (b) $V_-=0$.
    All frequencies and interaction strengths in units of $\pump$.
    }
    \label{fig:spectra}
\end{figure}

In \cref{fig:spectra}(a), we show the spectrum $P_{A}(\omega)$ of species $A$ as a function of the nonreciprocal coupling $V_-$ for $\delta = 0$.
The synchronized phase exhibits a single peak at zero frequency, while the traveling-wave phase is indicated by peaks in the spectrum at nonzero frequencies.
At the critical value of $V_-$ that separates the two phases, we find an exceptional point at which the two eigenvectors of the matrix $M$ become collinear.
The exceptional point marks the crossing of a nonreciprocal phase transition~\cite{Fruchart_2021}. 
The presence of two peaks signals the two traveling-wave states with opposite chirality.

The spectrum in \cref{fig:spectra}(a) is symmetric under $\omega \rightarrow - \omega$.
This is a consequence of \PTsymm ($\mathcal{L} = \mathcal{L}^*$).
To see this, we consider the following expression of the two-time correlations in the long-time limit, which depends on the steady-state density matrix $\rho_\mathrm{ss}$~\cite{Breuer},
\begin{equation}
\lim_{t\rightarrow \infty}
\langle{\sigma}_{a}^+({t+\tau}){\sigma}_{b}^-(t)\rangle
=
\Tr[\sigma^+_a e^{\mathcal{L}\tau} ( \sigma^-_b \rho_\mathrm{ss} )]
\, .
\end{equation}
When the Liouvillian is $\mathcal{PT}$ symmetric, the unique steady state is real valued, as we have shown in the previous section.
Therefore, the two-time correlations are also real valued and
\begin{equation}
    \mathcal{L}=\mathcal{L}^*
    \Rightarrow
    \abs{P_{a}(\omega)} = \abs{{P}_{a}(-\omega)}  \, .
    \label{eq:spectrum_symmetric}
\end{equation}
Thus, the \PTsymm enforces that the two traveling-wave states with opposite chirality occur with equal weight in the spectral density.

Another exceptional point occurs when explicitly breaking \PTsymm by $\delta$, see \cref{fig:spectra}(b), where we set $V_- = 0$.
For large detuning, the spectrum indicates an unsynchronized state in which $A$ oscillates at $+\delta/2$, while $B$ oscillates at $-\delta/2$ (not shown).
At the transition between synchronized and unsynchronized state, $M$ exhibits an exceptional point.
In this case, however, the spectrum is not symmetric under $\omega \rightarrow -\omega$, and the transition stems from breaking \PTsymm explicitly.

\begin{figure}
    \centering
    \includegraphics[width = 3.1in]{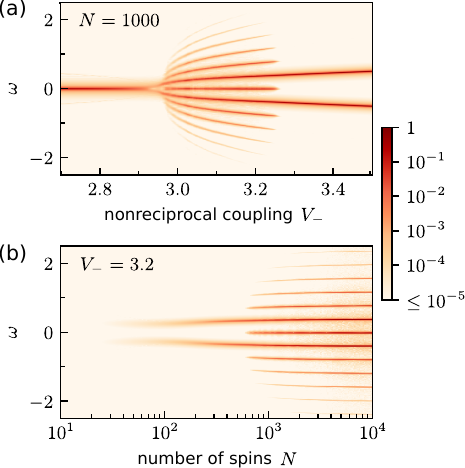}
    \caption{
    Spectrum $\abs{P_{A}(\omega)}$
    for finite-size systems
    shown on a logarithmic scale normalized to the maximum value.
    The spectrum is obtained by using \cref{eq:Mmatrix} together with the cumulant expansion.
    (a) Transition from the synchronized to the traveling-wave state via the modulated traveling-wave state for $N=1000$.
    In the modulated phase,
    higher harmonics appear in the spectrum.
    (b)
    Spectrum of the modulated traveling-wave state as a function of $N$ at $V_- = 3.2$.
    Parameters: $\delta=0$, $V_+ = V=2$. All frequencies and interaction strengths in units of $\pump$.
    }
    \label{fig:modphasespectrum}
\end{figure}

Finally, the dynamics of the modulated traveling-wave state are also observable in the spectrum of a finite-size system, see \cref{fig:modphasespectrum}(a).
Between the synchronized phase and the traveling-wave phase, a comb of frequencies opens up which signals a more intricate time-crystalline order.
The higher harmonics in the spectrum are consistent with the nonlinear evolution of the phases in the thermodynamic limit, see \cref{fig:fig1}(g).
The comb only becomes visible for $N \gtrsim 600$ spins as shown in \cref{fig:modphasespectrum}(b).
With increasing $N$ the comb structure becomes more pronounced.

\section{Quantum trajectories}
\label{s:qu_trajectories}

So far, we have analyzed quantities obtained by computing expectation values of operators using the density matrix.
The density matrix can be viewed as an ensemble description of quantum trajectories~\cite{Wiseman_2009}.
Each quantum trajectory is the evolution of the quantum system conditioned on the knowledge obtained by measurements.
In this section, we will study individual quantum trajectories.

Our model allows to measure (and thus obtain information about) the spin degrees of freedom by measuring the modes which mediate the interactions.
For simplicity, we focus on the case $V_+ = V$ while varying $V_-$ and consider the implementation depicted in \cref{fig:fig1}(c).
To model the measurement, we include in the master equation the coupling to mode $a_+$, which mediates reciprocal interactions
\begin{equation}
\begin{split}
    \dot \rho =
    &-i[\frac{\Omega}{2}(\amod^\dag S^- + \amod S^+) + H_0 + H_\mathrm{inter},\rho] + 
    \\
    &+\cavdec \mathcal{D}[\amod]\rho + \mathcal{L}_\mathrm{drive}\rho
    \equiv \mathcal{L}_\mathrm{m}\rho
    \, ,
    \end{split}
    \label{eq:master_cavity}
\end{equation}
where $S^\pm = S_{A}^\pm + S_{B}^\pm$ and $\kappa$ is the decay rate of mode $a_+$, which can be thought of as a lossy cavity.
In the limit $\cavdec \gg \Omega$, the cavity mode can be adiabatically eliminated, $\amod = -i\Omega S^-/\cavdec$, and \cref{eq:master_cavity} reduces to our master equation~\eqref{eq:master_spin} with $V = V_+ = Ng_+^2 = N \Omega^2 / \cavdec$.

The dynamics can be observed experimentally by standard homodyne and heterodyne detection techniques: the output of the cavity $\amod$ is mixed with a constant signal from a local oscillator with frequency $\omega_\meas$.
The evolution of the quantum state is then determined by the stochastic quantum master equation~\cite{Wiseman_2009}
\begin{equation}
\begin{split}
    \dot \rho_\meas =
    &\mathcal{L}_\mathrm{m}\rho_\meas +
    \frac{\mathrm{d}W}{\mathrm{d}t}
    \sqrt{\cavdec \xi}
    \left[e^{i\phi_\mathrm{m}(t)} ( \amod - \expval{\amod}_\meas ) \rho_\meas + \mathrm{H.c.} \right]
    \, ,
\end{split}
    \label{eq:master_stochastic}
\end{equation}
 where $\rho_\meas$ is the density matrix conditioned on the measurement outcome and 
$\expval{\amod}_\meas = \Tr[\amod \rho_\meas]$.
The measurement backaction is described by the random Wiener increment $\mathrm{d}W$, which follows a normal distribution with variance $\mathrm{d}W^2 = \mathrm{d}t$ and zero mean.
The detection efficiency $\xi$ can take values $\xi \in [0,1]$.
For $\xi=0$, no information about the quantum state is obtained, and \cref{eq:master_stochastic} reduces to \cref{eq:master_cavity}.
For simplicity, we focus on an ideal detector with $\xi = 1$.
The phase $\phi_\mathrm{m}$ determines how the cavity field $a_+$ is monitored.
When the phase is set constant (in the frame rotating with the cavity field $a_+$), the measurement is called \textit{homodyne} detection, and it probes the quadrature $a_+ e^{i\phi_\mathrm{m}} + a_+^\dag e^{-i\phi_\mathrm{m}}$.
When the phase increases with time as $\phi_\mathrm{m}(t) = \omega_\meas t$,
where $\omega_\meas$ is large compared to the timescale at which the dynamics occur, the measurement is called \textit{heterodyne} detection.
Thus, all possible quadratures are probed in quick succession, and the measurement becomes effectively isotropic.

To solve \cref{eq:master_stochastic}, we resort to a cumulant expansion~\cite{Zhang_2022} which we have already shown in \cref{s:finitesize} to describe well the full density matrix, see details in \cref{app:trajectories}.

\subsection{Heterodyne measurement}
\label{s:heterodyne}
\begin{figure*}
    \centering
    \includegraphics[width=\linewidth]{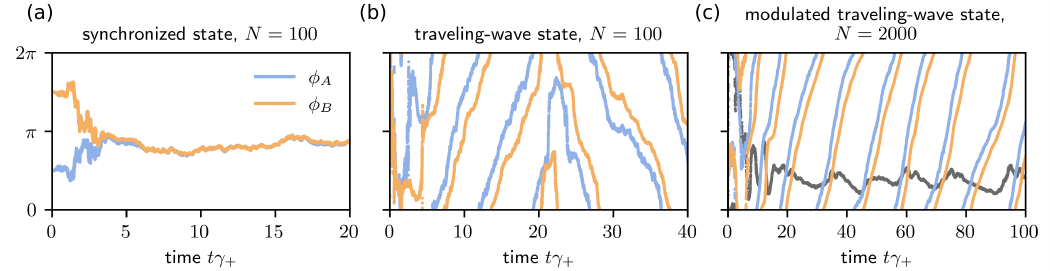}
    \caption{
    Time evolution under heterodyne detection of the phases of the spins.
    (a)
    Synchronized state for $V_- = V_+/2$ and $N=100$.
    The phases of spins $A$ and $B$ synchronize and a state that is static up to noise is obtained.
    (b)
    Traveling-wave state for $V_- = 2V_+$ and $N=100$.
    The phases of the spin species continuously oscillate while the phase difference remains constant.
    At around $t\gamma_+ = 22$, the chirality of the state switches.
    (c) Modulated traveling-wave states for $V_- = 1.6V_+$, here shown for $N = 2000$.
    The grey line shows the phase difference which is not static (compare to \cref{fig:fig1}(g)).
    Parameters:
    $\delta=0$,
    $\xi = 1$,      
    $\omega_\meas/\cavdec = 2\pi$,
    $\Omega/\cavdec = 1 / \sqrt{5N}$,
    $V_+ = V = 2\pump$
    }
    \label{fig:traj_sync_TW}
\end{figure*}

\Cref{fig:traj_sync_TW} shows quantum trajectories in the presence of heterodyne detection of synchronized, traveling-wave, and modulated traveling-wave states, respectively.
These results can be compared with the time evolution in the thermodynamic limit depicted in \cref{fig:fig1}(e-g).
In the synchronized state, see \cref{fig:traj_sync_TW}(a), the phases $\phi_a \equiv \arg(\expval{\sigma^+_a}_\meas)$ of the two spin species $a\in\{A,B\}$ assume nearly the same constant value after some transient evolution.
The fluctuations in the phases of the spins stem from the measurement noise affecting the cavity field, see \cref{eq:master_stochastic}, which in turn acts on the spin degrees of freedom due to the coupling between spins and cavity.
For any individual quantum trajectory, the $U(1)$-symmetry of the master equation~\eqref{eq:master_spin} is spontaneously broken (see also~\cite{Zhang_2022}).
Averaging over many trajectories recovers the density matrix description where $\expval{\sigma^+_a} = 0$ in the steady state.

In the quantum traveling-wave state shown in panel (b) of \cref{fig:traj_sync_TW}, the spins oscillate with one of two chiralities, i.e., positive or negative frequency, and the phase difference assumes values close to $\pm \pi/2$.
The continuous time-translation invariance of Eq.~\eqref{eq:master_spin} is spontaneously broken due to the measurement backaction (see also~\cite{Cabot_2023}).
On the other hand, the $U(1)$-symmetry is dynamically restored, i.e., no phase is preferred in the time average of a single trajectory~\cite{Fruchart_2021}.
Panel (c) of \cref{fig:traj_sync_TW} shows a modulated traveling-wave state.
In this case, we have set $N=2000$ since only for larger values of $N$, the additional dynamics occur, see~\cref{fig:modphasespectrum}(b).

The measurement backaction spontaneously breaks \PTsymm: at different times, one of the two possible traveling-wave states is assumed.
Furthermore, the fluctuations in the phases of the spins due to the measurement can cause chirality reversals, i.e., switches between the two traveling-wave states.
One such reversal occurs in the trajectory displayed in \cref{fig:traj_sync_TW}(b) at around $t\gamma_+ = 22$.
This is similar to the classical nonreciprocal Kuramoto model in which noise can induce chirality reversals~\cite{Fruchart_2021}.
Remarkably, here, the chirality reversals occur due to the measurement noise, which is unavoidable in quantum systems.

\begin{figure*}
    \centering
    \includegraphics[width=\linewidth]{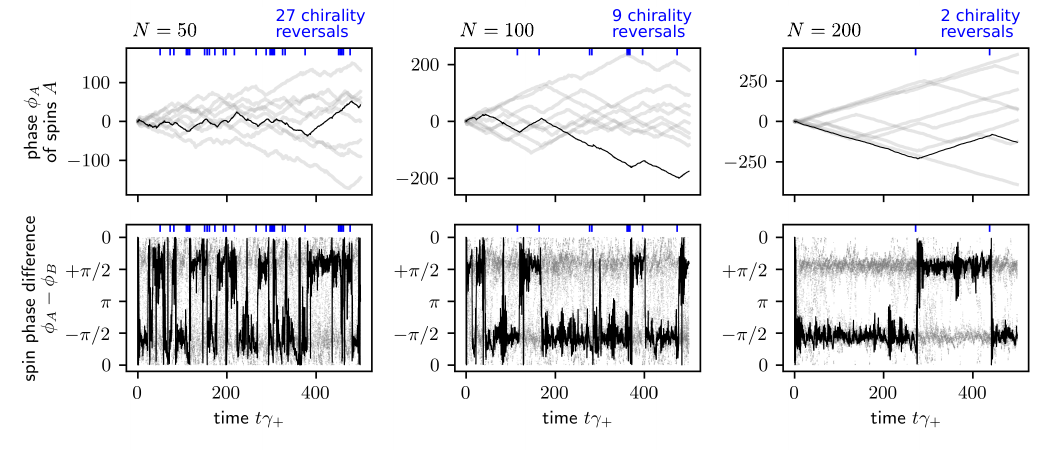}
    \caption{
    Time evolution under heterodyne detection of the phase of the spins of species $A$ (top row) and phase difference (bottom row) for different numbers of spins (columns).
    Several quantum trajectories are shown in grey, and one is highlighted in black.
    The top and bottom row in each column show the same quantum trajectories.
    In the top row, the phase is unwrapped, i.e., $2\pi$ is added or subtracted to the value of the phases such that the absolute difference between adjacent phase values never exceeds $\pi$. This allows to better distinguish periods of time of increasing and decreasing phase.
    Chirality reversals for $t\gamma_+>50$ are marked as blue ticks on the upper horizontal axes, respectively. 
    Parameters: same as \cref{fig:traj_sync_TW}(b) with $N$ specified for each column.
    }
    \label{fig:traj}
\end{figure*}

To analyze further the chirality reversals, we show the time evolution of the phase of spins of species $A$ in \cref{fig:traj} for several quantum trajectories (grey) with one highlighted in black.
Traveling-wave states with positive frequency are apparent as an increasing phase, while traveling-wave states with opposite chirality, i.e., negative frequency, are apparent as a decreasing phase.
The second row of \cref{fig:traj} displays the phase difference between spins of species $A$ and $B$.
The phase difference typically assumes values close to $\pm \pi/2$.
As expected from the dynamics in the thermodynamic limit, see \cref{fig:fig1}(e), the phase increases when the phase difference is close to $+\pi/2$ and decreases when  the phase difference is close to $-\pi/2$.
Also, the chirality reversals coincide with switches between the two possible values of the phase difference.
We highlight the chirality reversals as blue ticks on the upper horizontal axes.
To obtain them, the phase difference is time-averaged in a window of width $10/\pump$ (which corresponds roughly to one oscillation depending on $V_-$); a chirality reversal is then counted whenever the phase difference switches between a value closer to $\pi/2$ and a value closer to $-\pi/2$.

\begin{figure}
    \centering
     \includegraphics[width=3.4in]{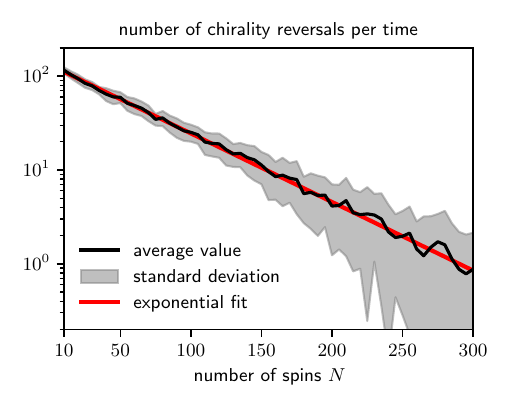}
    \caption{
    Number of chirality reversal as a function of $N$.
    The exponential fit $\propto \exp(-N/\delta N)$ yields $\delta N \approx 59$.
    The reversals are counted for 32 trajectories, each integrated for a total time of $10^3 / \pump$, of which the first $10^2/\pump$ are discarded as a transient.
    To obtain the number of reversals, the phase difference was averaged over a time window of $10/\pump$.
    Different values of the width of this window do not change the results qualitatively.
    The black line (grey area) shows the average (standard deviation) of $32$ trajectories.
    Parameters: same as \cref{fig:traj_sync_TW}(b).
    }
\label{fig:chiral_switches_exponential}
\end{figure}
In \cref{fig:chiral_switches_exponential}, we see that
the number of chirality reversals per fixed time decreases exponentially with the number of spins.
This is expected from classical models~\cite{Fruchart_2021}, and it is compatible with fluctuations whose strength decreases in proportion to $1/N$.
In the thermodynamic limit, no chirality reversals occur and the 
system remains in one of the two traveling-wave states.

Finally, we emphasize that these dynamics are observable in an experiment.
The heterodyne signal which is experimentally accessible is~\cite{Wiseman_2009}
\begin{equation}
    J = 2 \sqrt{\xi \kappa} \Re[e^{i\phi_\meas(t)} \expval{a}_\meas] + \mathrm{d}W/\mathrm{d}t
\end{equation}
Mixing this signal with $\cos(\phi_\meas(t))$ and $\sin(\phi_\meas(t))$ and time averaging allows to obtain the cavity quadratures $\Re[\expval{a_+}_\meas] $ and $\Im[\expval{a_+}_\meas]$ and thus the cavity phase. 
Since the cavity is to a good approximation related to the spins by $a_+ \approx -i\Omega S^-/\kappa$, the two traveling-wave states can be detected by an increasing or decreasing cavity phase.
Also, while we show results for unit measurement efficiency $\xi=1$, we tested that the traveling-wave states persist for efficiencies as low as $\xi \approx 0.1$.

\subsection{Homodyne measurement}
\label{s:homodyne}

In this section, we show that the measurement backaction can qualitatively influence the traveling-wave state.
Specifically, we investigate the influence of a homodyne detection of the quadrature $a_+ + a_+^\dag$, setting $\phi_\meas = 0$.
Quantum trajectories in the presence of homodyne detection reveal a traveling-wave state similar to that shown in \cref{fig:traj_sync_TW} for the case of heterodyne detection.
Nevertheless, as we will now show, there are qualitative differences between the traveling-wave states obtained via homodyne and heterodyne detection.

The measurement of a single cavity quadrature continuously projects the state of the cavity in states with well-defined expectation value along the $a_+ + a_+^\dag$ quadrature.
Due to the coupling between spins and cavity, the measurement backaction on $a_+$ also affects the spin degrees of freedom.
Given the adiabatic relation $a_+\propto iS^-$, the collective spin is effectively measured in the quadrature $S^y = i(S^- - S^+)$.
Furthermore, the phase difference of spins of species $A$ and $B$ is approximately $\pm \pi/2$ (for traveling-wave states) and $S^- = S^-_A + S^-_B$;
consequently, the spins $A$ and $B$ are effectively measured in quadratures rotated by $\pm \pi/4$ relative to the $y$-quadrature.
Since the sign of the phase shift depends on the chirality of the traveling-wave state, the measurement backaction on the spins depends on the chirality.
Therefore, the spins tend to assume states where the quadratures along these directions are well defined, while noise is enhanced along quadratures rotated by 90°.

\begin{figure}
    \centering
    \includegraphics[width=3.4in]{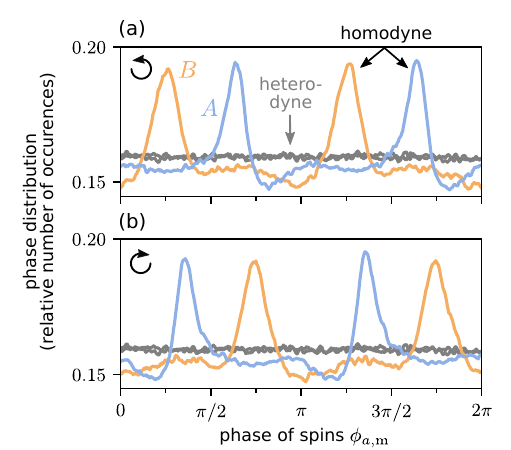}
    \caption{
    Phase distribution under heterodyne (grey, species $A$ and $B$) and under homodyne detection (blue: species $A$, orange: species $B$).
    The phase distributions are obtained by counting the occurrences of the phases of the spins averaged over time and multiple trajectories.
    Clearly, the phase distributions differ for homodyne and heterodyne measurements.
    Also, we postselect on trajectories which show a left chirality in panel (a) and a right chirality in panel (b).
    Parameters: same as \cref{fig:traj_sync_TW}(b) with $N=500$ (for homodyne measurements, we set $\omega_\meas=0$).
    }
    \label{fig:homodyne}
\end{figure}

The expected behavior is confirmed in \cref{fig:homodyne}, which displays the phase distribution of species $A$ and $B$.
The phase distributions quantify the likelihood of the spins to assume a particular phase value, averaged over time and multiple trajectories.
We have postselected states with left and right chirality, and show the results for the two chiralities in Panels (a) and (b), respectively.
For the heterodyne case, the measurement backaction is isotropic.
For the homodyne case, on the other hand, the backaction on the state of the spins $A$ and $B$ depends on their phases $\phi_{a,\meas} = \arg(\expval{\sigma^+_{a}}_\meas)$.
The backaction results in peaks of the phase distribution which occur at phases of approximately $\pi/2 \pm \pi/4$ and  $-\pi/2 \pm \pi/4$ as expected from the discussion above.
The dependence of the distribution on the chirality is evident by comparing Panels (a) and (b).
Importantly, \cref{fig:homodyne} clearly shows that the measurement backaction qualitatively influences the traveling-wave states.

It will be interesting to investigate homodyne detection for smaller ensemble sizes, where the measurement noise is stronger.
However, we find that the cumulant expansion fails to approximate well the state of the system for $N\lesssim 200$, which hints at the presence of non-Gaussian correlations.
We expect to find more qualitative differences between quantum trajectories in the presence of heterodyne and homodyne detection at these smaller values of $N$.
Furthermore, the scaling of the number of chirality reversals may depend on the measurement.

\section{Implications for nonreciprocal  quantum many-body systems}
\label{s:general_discussion}

In the previous sections, we have  analyzed a specific model to show how to engineer nonreciprocal interactions in a quantum many-body system and how to identify their observable consequences such as dynamical traveling wave states.
While some aspects of our analysis are specific to the model considered, our results have implications for a broader class of quantum models featuring nonreciprocal interactions. In this section we highlight what we believe are the general features of our analysis 
as well as discuss possible directions to extend our framework. 

\begin{figure*}
    \includegraphics[width = 7in]{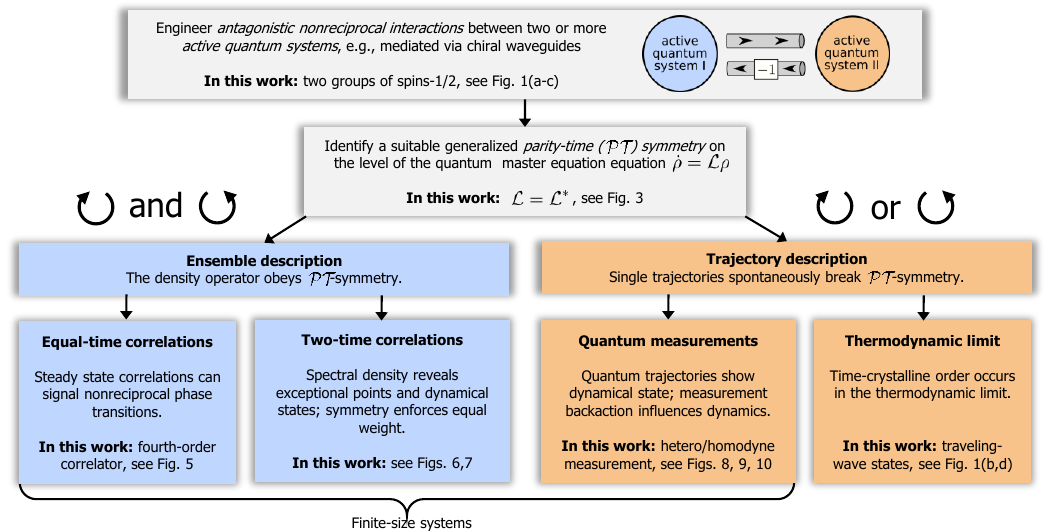}
    \caption{
    Schematic diagram for the analysis of nonreciprocal phase transitions in active quantum systems.
    }
    \label{fig:schematic}
\end{figure*}

In Fig.~\ref{fig:schematic} we show a flowchart illustrating the key conceptual ingredients in the analysis of a nonreciprocal quantum many-body system. The main features are:

\paragraph{Engineering antagonistic nonreciprocal interactions.}
 
Nonreciprocal interactions are not native to quantum systems and must be engineered instead.
Chiral waveguides, in combination with phase shifts along their path, can mediate tuneable antagonistic nonreciprocal interactions between two subsystems $A$ and $B$, see \cref{fig:fig1}(b,c).
While in this work we focused on the case where both $A$ and $B$ are a collection of quantum spins-1/2, our framework is in principle agnostic to the nature of the subsystems.
The reduced master equation~\eqref{eq:master_spin} can be employed to model antagonistic interactions
among other subsystems, e.g., comprising constituents with a larger local Hilbert space.
Additionally, the framework allows to study nonreciprocal couplings in a network comprising more than two subsystems, exploring the role of community effects and network topology.

\paragraph{Active quantum systems.}
Nonreciprocal interactions among agents are enabled  by the out-of-equilibrium character of active matter~\cite{Fruchart_2021, Bowick_2022}.
We showed that, analogous to classical active matter, a quantum system provided with a source of energy and thus driven out of equilibrium, becomes active and can feature nonreciprocal interactions.
While in our model a quantum many-body system is promoted to an active quantum system by the local incoherent drive of~\cref{eq:master_L_drive}, other mechanisms may be conceived, such as saturated gain as in lasers or even coherent light-matter interactions beyond the rotating-wave approximation~\cite{Kirton_2019,Chiacchio_2023}.

\paragraph{The role of \PTsymm.}
\label{implication:role_of_pt}
Nonreciprocal phase transitions are \PTsymm breaking transitions~\cite{Fruchart_2021}. 
Since open quantum systems are described in terms of a quantum master equation~\cite{Breuer}, or equivalently, a Liouvillian $\mathcal{L}$,
\PTsymm should be identified at the level of $\mathcal{L}$. 
In our model, the implementation based on chiral waveguides allows to identify \PTsymm with complex conjugation of the Liouvillian $\mathcal{L} = \mathcal{L^*}$. We expect this notion of \PTsymm to be relevant  beyond the model considered here.
While our definition has the twofold advantage of being physically motivated and recovering (in the thermodynamic limit) the generalized \PTsymm used in \cite{Fruchart_2021}, parity-time symmetry is a broad concept~\cite{Bender_2023}
and several instances have been discussed in the literature~\cite{Prosen_2012,Huber_2020,Sa2023,Huybrechts_2020,Nakanishi_2022}.

\paragraph{Symmetry constraints in ensemble description.}
For any finite-size system, the  master equation has typically a unique time-independent steady state.
Traveling-wave states are therefore not directly visible from the steady-state density matrix.
Nevertheless, signatures of the time-crystal phase can still be identified at the ensemble level, by inspecting either  equal-time or two-time correlation functions (or equivalently the spectral density). 
In our model, the spectral density satisfies $\abs{P(\omega)} = \abs{P(-\omega)}$, which implies that the two traveling-wave states with opposite chirality, i.e., frequency, have equal weight. 
Following from symmetry arguments, we expect this to be a general feature of $\mathcal{PT}$ symmetric Liouvillians. 

\paragraph{Spontaneous \PTsymm breaking in the thermodynamic limit.}
While the ensemble description obeys \PTsymm, it is nonetheless possible to observe the spontaneous breaking of \PTsymm in two qualitatively different ways.
One way is by taking the thermodynamic limit, where
\PTsymm may be spontaneously broken (the system assumes one of two possible chiralities) in conjunction with the breaking of continuous time translational symmetry. 
This establishes that a quantum model can feature time-crystalline states from dynamical frustration~\cite{Hanai_2024}, i.e., nonreciprocal interactions.

\paragraph{Quantum trajectories and measurement backaction.}
A second way to observe spontaneous \PTsymm breaking, for any finite system size, 
is through measurement backaction.
In quantum systems, the effects of a measurement are described by unraveling the dynamics into quantum trajectories, where each trajectory represents a possible realization of the system’s evolution under continuous observation.
Individual quantum trajectories reveal the oscillatory behavior characteristic of the time-crystal and the associated symmetry breaking.
Additionally, the measurement backaction, i.e., the influence of the chosen measurement on the quantum state, qualitatively alters the dynamics of quantum traveling-wave states. 

\paragraph{The role of decoherence.}

In classical active matter, noise or disorder is crucial for perturbing potential equilibrium states, thereby enabling the emergence of traveling-wave states~\cite{Fruchart_2021}.
In our model, decoherence takes up this role:
by destabilizing any state with maximal coherence, it allows for persistent traveling-wave states.
In any quantum system, the coupling to an environment unavoidably leads to decoherence.
Therefore, it can be expected that decoherence acts to stabilize dynamical states in other quantum models featuring nonreciprocal interactions.

\section{Conclusions}

Nonreciprocal interactions in active matter lead to exciting features such as a new class of critical phenomena and phase transitions.
Since the laws of nature are fundamentally quantum mechanical, these features have to emerge from an underlying microscopic quantum theory.
Our work shows how this is possible in the paradigmatic model of synchronization.
To this end, we have described a way for active quantum agents to interact nonreciprocally.
Each constituent is an active quantum spin that is incoherently driven.
The phases of one species of spins are attracted to those of the other species, which in contrast are repelled.
We have discussed how these antagonistic quantum many-body interactions result in traveling-wave states that arise via a nonreciprocal phase transition.
We have demonstrated the connection of this kind of phase transition to the  spontaneous breaking of parity-time ($\mathcal{PT}$)-symmetry which we have motivated from physical grounds and formulated as an invariance of the Liouvillian under complex conjugation.

The model can be implemented in current experimental settings and therefore offers a new platform to investigate the importance of nonreciprocal interactions in the quantum domain.
Furthermore, we have demonstrated that the traveling-wave state remains observable for finite-size systems.
On the level of the density matrix, fourth-order correlations and exceptional points in the spectrum provide ways to detect traveling-wave states.
Most intriguingly, standard heterodyne detection measurements allow to observe quantum trajectories featuring traveling-wave states that spontaneously break \PTsymm.

Our results are significant beyond the particular model considered here as emphasized in the previous section.
Chiral waveguides and phase shifts can be used to engineer nonreciprocal interactions among arbitrary active quantum systems.
We  have highlighted the importance of the formulation of \PTsymm on the level of the master equation.
It constrains the steady state of the system and is spontaneously broken in the
thermodynamic limit and in quantum trajectories.
The measurement with which the quantum trajectories are obtained acts back on the quantum system and influences the dynamics.
Furthermore, decoherence stabilizes a time-crystal state and thus takes the role of 
noise or disorder which stabilize the dynamical state in classical active matter~\cite{Fruchart_2021}.
We expect these insights to be useful in future studies of nonreciprocity in quantum systems.

Antagonistic interactions of active matter in the quantum domain are a novel research direction with exciting further questions.
First of all, it will be interesting to study the influence of nonreciprocity in other systems using our framework to engineer nonreciprocal interactions.
Furthermore, it is intriguing to study various symmetries.
While we have identified both \PTsymm  and \Usymm, a different model showed a \PTsymm together with a $\mathbb{Z}_2$-symmetry~\cite{Chiacchio_2023}.
It would also be interesting to find an instance of strong \PTsymm and see if it can be spontaneously broken. 
Recently, a close connection between geometric and dynamical frustration has been pointed out~\cite{Hanai_2024}.
Geometric frustration is known to lead to exciting phases of quantum matter, such as quantum spin liquids, and it will be interesting to explore the consequences of nonreciprocal dynamical frustration in quantum systems.

The field of active quantum matter is just emerging~\cite{Khasseh_2023}.
Given the relevance of antagonistic interactions in classical active matter, it is important to consider similar interactions of quantum constituents.
Our work proposes a way to do so and highlights general features of nonreciprocal interactions in active quantum matter.

\begin{acknowledgments}
We would like to thank Philipp Treutlein for 
insightful discussions about the experimental implementation.
We also thank
Federico Carollo, Andreas Nunnenkamp, and Marco Schir\`o for useful discussions.
We acknowledge financial support from the Swiss National Science Foundation (T.N.\ and C.B.: grant No.\ 200020 200481, M.B.: grant No.\ PCEFP2\_194268).
\end{acknowledgments}

\appendix

\section{General complex coherent coupling}
\label{app:general_Vm}

\begin{figure}
    \centering
    \includegraphics[width = 1.7in]{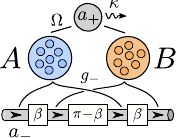}
    \caption{Setup to implement complex-valued nonreciprocal coupling $V_-$.
    The cavity mode $a_+$ mediates reciprocal interactions.
    The phase shifters in the unidirectional chiral mode $a_-$ transform $a_- \rightarrow \exp(i\beta)a_- $ or $a_- \rightarrow \exp(i(\pi-\beta))a_-$.
    }
    \label{fig:general_Vm}
\end{figure}

We describe how a complex-valued nonreciprocal coupling $V_-$ can be implemented.
The setup is shown in \cref{fig:general_Vm}.
Following \cite{Karg_2019}, 
the unidirectional waveguide with mode $a_-$ mediates the coherent interactions of our master equation~\eqref{eq:master_spin} with
\begin{equation}
    V_-/N = 2 g_-^2 e^{i\beta} \, .
\end{equation}
The cavity mode $a_+$ mediates dissipative interspecies and intraspecies interactions with
\begin{equation}
    V_+/N = V/N = 2 g_+^2 \, .
\end{equation}
For simplicity, we focus on the case where there are no losses in the unidirectional waveguide.

Under time-reversal the chirality of the unidirectional waveguide is reversed and the phase shifts $\beta$ and $\pi-\beta$ also change sign.
The setup shown in \cref{fig:general_Vm} is only invariant under the \PTtrans for $\beta \in \{0,\pi\}$ which corresponds to real-valued $V_-$.
This confirms that the imaginary part of $V_-$ explicitly breaks \PTsymm.

In the thermodynamic limit, we obtain the mean-field equations
\begin{equation}
\begin{aligned}
  \frac{\mathrm{d}}{\mathrm{d}t} {s}_{A}^+
  =& [(-\pump + i \delta) s^+_{A}
  +V  s^+_{A} s^z_{A} 
  +(V_+ - V_-^*) s^+_B    s^z_{A}] /2
\, , \\
  \frac{\mathrm{d}}{\mathrm{d}t} {s}_{B}^+
  =&[(-\pump - i \delta) s^+_{B}
  +V  s^+_{B} s^z_{B} 
  +(V_+ + V_-) s^+_A    s^z_{B} ]/2
\, , \\
  \frac{\mathrm{d}}{\mathrm{d}t} {s}_{A}^z
  =&
  \pump\left( 1- {s}_{A}^z \right)
  -2V {s}_{A}^+{s}_{A}^- 
  -2\Re[(V_+ - V_-^*){s}_{A}^-{s}_{B}^+]
 \\
  \frac{\mathrm{d}}{\mathrm{d}t} {s}_{B}^z
  =&
  \pump\left( 1- {s}_{B}^z \right)
  -2V {s}_{B}^+{s}_{B}^- 
  -2\Re[(V_+ + V_-){s}_{A}^+ {s}_{B}^-]
  \, ,
\end{aligned}
\end{equation}
which reduce to \cref{eq:MF} when $V_-$ is real valued.

\section{{Phase diagrams}}
\label{app:phase_diagrams}

\subsection{Transition between inhomogeneous and synchronized state}
\label{app:trivial_to_synch}

\begin{figure}
    \centering
    \includegraphics[width = 3.4in]{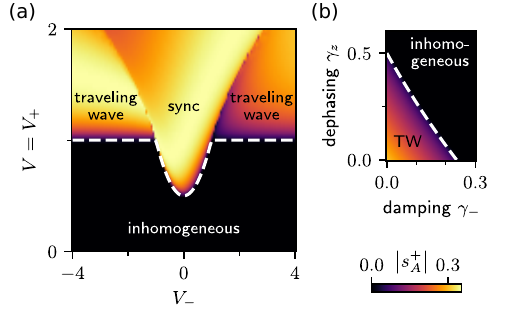}
    \caption{Transition from inhomogeneous to synchronized states in the thermodynamic limit indicated by the absolute value of the coherence $\abs{s^+_A}$ (similar results are obtained for $\abs{s^+_B}$).
    (a)
    Transition from inhomogeneous to synchronized states by increasing $V=V_+$ as a function of $V_-$.
    There is both a transition to the static synchronized state and to the dynamic traveling-wave state.
    The white dashed line is obtained from a stability analysis of the inhomogeneous state,
    $V = V_+ = \mathrm{min}(1,(1+V_-^2)/2)$, a special case of \cref{eq:app_stability} in the absence of damping and dephasing.
    All interaction strengths in units of $\pump$ and $\delta=0$.
    (b) 
    Transition of the traveling-wave (TW) state to the inhomomgeneous state by increasing damping and decay.
    The white dashed line is $\Gamma = Vz_0$ following \cref{eq:app_stability}.
    Parameters: $\delta=0$, $V_- = 2V_+ = 2V = 4\pump$.
    Damping and dephasing strengths are given in units of $\pump$.
    }
    \label{fig:app_trivial_to_synch}
\end{figure}

Throughout the main text we focus on the case where $V/\pump$ is large enough such that all spins of each species are synchronized.
\Cref{fig:app_trivial_to_synch}(a) displays the transition from an inhomogeneous state to either the static synchronized state or the dynamical traveling-wave state by increasing the dissipative coupling strengths.
The transition is clearly indicated by the absolute value of the mean coherence $\abs{s^+_A}$ (or equivalently $\abs{s^+_B}$, not shown), which is zero in the inhomogeneous regime.

Damping and dephasing of the spins can also induce a transition between synchronized and inhomogeneous states.
Importantly, however, the traveling-wave state (and the other synchronized states) possesses some degree of robustness with respect to damping and dephasing. 
To show this, we include damping at rate $\gamma_-$ and dephasing at rate $\gamma_z$
by adding to the master equation \eqref{eq:master_spin} the terms
\begin{equation}
\begin{split}
    \sum_{a\in\{A,B\}}
        \sum_{i=1}^N 
        \gamma_- \mathcal{D}[{\sigma}_{a,i}^-]\rho +
        \gamma_z \mathcal{D}[{\sigma}_{a,i}^z]\rho\,.
\end{split}
\end{equation}
In the thermodynamic limit, the equations \cref{eq:MF} become (setting $\delta = 0$ for simplicity),
\begin{equation}
\begin{split}
  \frac{\mathrm{d}}{\mathrm{d}t} {s}_{A}^+
  =&[-(\pump +\gamma_- +2\gamma_z) s^+_{A}
  +V  s^+_{A} s^z_{A} 
  +V_{BA} s^+_B    s^z_{A}] /2
\, , \\
  \frac{\mathrm{d}}{\mathrm{d}t} {s}_{B}^+
  =&[-(\pump +\gamma_- +2\gamma_z) s^+_{B}
  +V  s^+_{B} s^z_{B} 
  +V_{AB} s^+_A    s^z_{B}] /2
\, ,
\\
  \frac{\mathrm{d}}{\mathrm{d}t} {s}_{A}^z
  =&
  \pump\left( 1- {s}_{A}^z \right)
  -
  \gamma_- \left( 1+ {s}_{A}^z \right)
  \\
  &-2V {s}_{A}^+{s}_{A}^- 
  -2V_{BA}\Re[{s}_{A}^+{s}_{B}^-]
  \, ,
 \\
  \frac{\mathrm{d}}{\mathrm{d}t} {s}_{B}^z 
  =&
  \pump\left( 1- {s}_{B}^z \right)
  -
  \gamma_- \left( 1+ {s}_{B}^z \right)
  \\
  &-2V {s}_{B}^+{s}_{B}^- 
  -2V_{AB}\Re[{s}_{A}^+{s}_{B}^-]
  \, .
\end{split}
\label{eq:app_mf_damping_dephasing}
\end{equation}
We find that the incoherent steady state
\begin{equation}
    s^+_{A,B} = 0 \, , \quad s^z_{A,B} = z_0 \equiv \frac{\pump - \gamma_-}{\pump+\gamma_-}
\end{equation}
becomes unstable when
\begin{equation}
\Gamma <
    \begin{cases} 
        z_0 V & \text{for } \abs{V_-} > \abs{V_+} \, , \\
        z_0 \left( V + \sqrt{V_+^2 - V_-^2} \right) & \text{for } \abs{V_-} < \abs{V_+} \, ,
    \end{cases}
    \label{eq:app_stability}
\end{equation}
where we defined the total decoherence rate $\Gamma = \pump + \gamma_- + 2\gamma_z$.

We show the influence of damping and dephasing on the coherence $\abs{s^+_A}$ of a traveling-wave state in \cref{fig:app_trivial_to_synch}(b).
Importantly, the state also retains its dynamics throughout the region where $\abs{s^+_A}$ is finite (not shown).
We conclude that time-crystalline traveling-wave states are robust to damping and dephasing.

\subsection{Amplitude and phase difference}
\label{app:phase_diagrams_data}
\begin{figure}
    \centering
    \includegraphics[width = 3.4in]{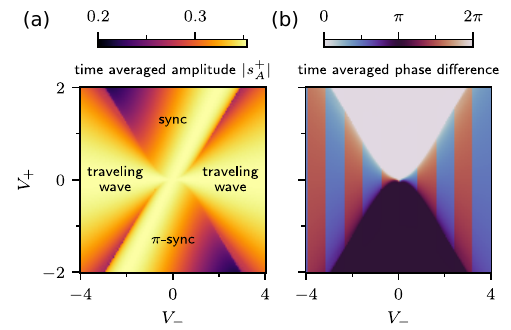}
    \caption{
    Phase diagram in thermodynamic limit.
    (a) Time-averaged amplitude $\abs{s^+_A}$ in the long-time limit as a function of $V_-$ and $V_+$.
    (b) Time-averaged phase difference $\arg{s^+_A/s^+_B}$ in the long-time limit.
    The initial conditions are chosen such that either of the two traveling-wave states with phase difference $\pm \pi/2$ is obtained in alternating vertical stripes.
    Parameters: $\delta=0$, $V=2$. All coupling strengths in units of $\pump$.
    }
\label{fig:app_phase_difference_plot}
\end{figure}

The data used to produce \cref{fig:fig1}(d) is shown in  \cref{fig:app_phase_difference_plot}.
Panel (a) shows the time-averaged amplitude $\abs{s^+_A}$ in the long-time limit as a function of $V_-$ and $V_+$.
Panel (b) shows the time-averaged phase difference.
In both panels, one can identify boundaries between the regions of static ($\pi$-)synchronized states and the regions of dynamical traveling-wave states.
The modulated traveling-wave state is not apparent in \cref{fig:app_phase_difference_plot}, since the quantities are time averaged.

We find numerically that the phase difference shown in \cref{fig:app_phase_difference_plot}(b) is precisely $\pm \pi/2$ for $V_+ = 0$.
As $V_+$ increases and the spins approach the synchronized phase, the phase difference continuously shifts towards $0$.
Conversely, for negative $V_-$, as the spins approach the $\pi$-synchronized phase, the phase difference shifts towards $\pi$.

\section{Finite-size calculations of the steady state}
\label{app:finite_size_calculations}

To solve for the long-time limit of the full quantum master equation~\eqref{eq:master_spin},
we exploit the permutational symmetry, i.e., the fact that all two-level systems within each group are identical.
This reduces the complexity from $4^{2N}$ to $(N^2)^3$, where $N$ is the number of spins per species.
To find the steady state, we compute the eigenstate of the Liouvillian corresponding to the zero eigenvalue.
Numerics are performed using PIQS~\cite{Shammah_2018}.

For larger $N$, this approach becomes unfeasible such that we resort to an approximation.
We systematically include correlations by using cumulant expansions~\cite{cumulants_kubo} to second and fourth order, where third-order or fifth-order cumulants are set to zero, respectively.
For instance,
\begin{equation}
    \begin{split}
    \langle o_1o_2o_3 \rangle_c ={}& \langle o_1o_2o_3\rangle  -\langle o_1o_2\rangle\langle o_3\rangle - \langle o_1o_3\rangle\langle o_2\rangle
\\
&- \langle o_1\rangle\langle o_2o_3\rangle + 2\langle o_1\rangle\langle o_2\rangle\langle o_3\rangle
= 0
\, .
\end{split}
\label{eq:app_cumulant}
\end{equation}
Furthermore, we exploit the \Usymm of the master equation~\eqref{eq:master_spin} to set averages such as $\expval{\sigma^+_{a,i}}=0$ or $\langle \sigma^+_{a,i} \sigma^z_{b,j} \rangle=0$ to zero as they vanish in the long-time limit.
Additionally, we invoke the permutational invariance to set all spins within each species equal, e.g.,
$\langle{\sigma}_{A,i}^+{\sigma}_{A,j}^-\rangle =\langle{\sigma}_{A}^+{\sigma}_{A}^-\rangle$ for all $i \neq j$
or
$\langle{\sigma}_{A,i}^+{\sigma}_{B,j}^-\rangle =\langle{\sigma}_{A}^+{\sigma}_{B}^-\rangle$ for all $i$ and $j$.
In the second-order expansion, we thus obtain a closed set of differential equations for
$\expval{\sigma^z_{a}}$,
$\langle{\sigma}_{a}^+{\sigma}_{a}^-\rangle$,
$\langle{\sigma}_{A}^+{\sigma}_{B}^-\rangle$,
and
$\langle{\sigma}_{a}^z{\sigma}_{a}^z\rangle$,
and
$\langle{\sigma}_{A}^z{\sigma}_{B}^z\rangle$.
{
\allowdisplaybreaks
\begin{align*}
\frac{\mathrm{d}}{\mathrm{d}t}s_{A}^z
  =&-V\left(s_{A}^z+1\right)/N
  +\pump\left(1-s_{A}^z\right)
  \\
  &{}-2V \frac{N-1}{N}\langle{\sigma}_{A}^+{\sigma}_{A}^-\rangle 
  -2 V_{BA} \Re[\expval{\sigma^+_A\sigma^-_B}]
\, , \\
\frac{\mathrm{d}}{\mathrm{d}t}s_{B}^z
  =&-V\left(s_{B}^z+1\right)/N
  +\pump\left(1-s_{B}^z\right)
  \\
  &{}-2V \frac{N-1}{N}\langle{\sigma}_{B}^+{\sigma}_{B}^-\rangle
  -2 V_{AB}  \Re[\expval{\sigma^+_A\sigma^-_B}]
\, , \\
\frac{\mathrm{d}}{\mathrm{d}t}\langle{\sigma}_{A}^+{\sigma}_{A}^-\rangle
=&-(\pump + \frac{V}{N})\langle{\sigma}_{A}^+{\sigma}_{A}^-\rangle+\frac{V}{2N}
\left(\langle{\sigma}_{A}^z{\sigma}_{A}^z\rangle
  +s^z_A\right)\nonumber\\
&{}+Vs^z_A(N-2)/N
\langle{\sigma}_{A}^+{\sigma}_{A}^-\rangle\nonumber\\
&{}+
V_{BA} s_{A}^z \Re[\expval{\sigma^+_A\sigma^-_B}]
  \,, \\
\frac{\mathrm{d}}{\mathrm{d}t}\langle{\sigma}_{B}^+{\sigma}_{B}^-\rangle
=&-(\pump + \frac{V}{N})\langle{\sigma}_{B}^+{\sigma}_{B}^-\rangle+\frac{V}{2N}
\left(\langle{\sigma}_{B}^z{\sigma}_{B}^z\rangle
  +s^z_B\right)\nonumber\\
&{}+Vs^z_B(N-2)/N
\langle{\sigma}_{B}^+{\sigma}_{B}^-\rangle\nonumber\\
&{}+
V_{AB} s_{B}^z \Re[\expval{\sigma^+_A\sigma^-_B}]
  \,, \\
\frac{\mathrm{d}}{\mathrm{d}t}\langle{\sigma}_{A}^+{\sigma}_{B}^-\rangle
=&-(\pump +V/N-i\delta)\langle{\sigma}_{A}^+{\sigma}_{B}^-\rangle
\\
&+V\frac{N-1}{2N}\left(s^z_A +  s^z_B \right)\expval{\sigma^+_{A} \sigma^-_{B}}
\\
& + V_{AB}(s^z_B+\expval{\sigma^z_A\sigma^z_B})/4N
\\
& + V_{BA}(s^z_A+\expval{\sigma^z_A\sigma^z_B})/4N
\\
& + \frac{N-1}{2N} \left(
V_{AB} s^z_B \expval{\sigma^+_{A} \sigma^-_{A}} +
V_{BA} s^z_A \expval{\sigma^+_{B} \sigma^-_{B}}
\right)
\\
\frac{\mathrm{d}}{\mathrm{d}t}\langle{\sigma}_{A}^z{\sigma}_{A}^z\rangle
=&{}
2s_A^z(\pump - V/N) -2 \langle{\sigma}_{A}^z{\sigma}_{A}^z\rangle(\pump +V/N)
\\
&- 4 V_{BA}s^z_A \Re[\expval{\sigma^+_A\sigma^-_B}]
\\
&+V \left( 4\expval{\sigma^+_{A}\sigma^-_{A}}  - 4(N-2) s^z_A \expval{\sigma^+_{A}\sigma^-_{A}}\right)/N
\\
\frac{\mathrm{d}}{\mathrm{d}t}\langle{\sigma}_{B}^z{\sigma}_{B}^z\rangle
=&{}
2s_B^z(\pump - V/N) -2 \langle{\sigma}_{B}^z{\sigma}_{B}^z\rangle(\pump +V/N)
\\
&- 4 V_{AB}s^z_B \Re[\expval{\sigma^+_A\sigma^-_B}]
\\
&+V \left( 4\expval{\sigma^+_{B}\sigma^-_{B}}  - 4(N-2) s^z_B \expval{\sigma^+_{B}\sigma^-_{B}}\right)/N
\\
\frac{\mathrm{d}}{\mathrm{d}t}\langle{\sigma}_{A}^z{\sigma}_{B}^z\rangle
=&{}
\pump (s_A^z+s_B^z -2\langle{\sigma}_{A}^z{\sigma}_{B}^z\rangle)
\\
&
+ 2\frac{N-1}{N}V_-(s^z_B-s^z_A)\Re[\expval{\sigma^+_A\sigma^-_B}]
\\
&-2\frac{N-1}{N}V_+(s^z_A+s^z_B)\Re[\expval{\sigma^+_A\sigma^-_B}]
\\
&- 2\frac{N-1}{N} V (s^z_B\expval{\sigma^+_A\sigma^-_A} + s^z_A\expval{\sigma^+_B\sigma^-_B}) 
\\
&-V(s^z_A+s^z_B + 2\expval{\sigma^z_A\sigma^z_B})/N
\\
&+4V_+\Re[\expval{\sigma^+_A\sigma^-_B}]/N
\, .
\end{align*}
}
In the limit $N \rightarrow \infty$, they are equivalent to the mean-field equations.
For the fourth-order expansion, we use the julia package QuantumCumulants.jl~\cite{Plankensteiner2022quantumcumulantsjl}.

\section{Two-time correlations}
\label{app:twotime}

The two-time correlations evolve according to
\begin{equation}
\frac{\mathrm{d}}{\mathrm{d}\tau}
\begin{pmatrix}
\langle{\sigma}_{A}^+(t+\tau){\sigma}_{B}^-(t)\rangle\\
\langle{\sigma}_{B}^+(t+\tau){\sigma}_{B}^-(t)\rangle
\end{pmatrix}=
M
\begin{pmatrix}
\langle{\sigma}_{A}^+(t+\tau){\sigma}_{B}^-(t)\rangle\\
\langle{\sigma}_{B}^+(t+\tau){\sigma}_{B}^-(t)\rangle
\end{pmatrix},
\label{eq:app_2timecorrs_evolution1}
\end{equation}
and 
\begin{equation}
\frac{\mathrm{d}}{\mathrm{d}\tau}
\begin{pmatrix}
\langle{\sigma}_{A}^+(t+\tau){\sigma}_{A}^-(t)\rangle\\
\langle{\sigma}_{B}^+(t+\tau){\sigma}_{A}^-(t)\rangle
\end{pmatrix}=
M
\begin{pmatrix}
\langle{\sigma}_{A}^+(t+\tau){\sigma}_{A}^-(t)\rangle\\
\langle{\sigma}_{B}^+(t+\tau){\sigma}_{A}^-(t)\rangle
\end{pmatrix},
\label{eq:app_2timecorrs_evolution2}
\end{equation}
with
\begin{equation}
M= \frac{1}{2}
    \begin{pmatrix}
        -\gamma_A +i\delta + V s^z_A  &  V_{BA}  s^z_A  \\
        V_{AB} s^z_B   & -\gamma_B -i\delta  + V s^z_B 
    \end{pmatrix}
    \, ,
    \label{eq:app_Mmatrix}
\end{equation}
where the effective decoherence rate $\gamma_{A,B} = \pump + V{(1+s^z_{A,B})}/N$ depends on the number of spins $N$.
The populations $s^z_a$ that enter $M$ are to be evaluated at time $t+\tau$.
To derive Eqs.~\eqref{eq:app_2timecorrs_evolution1} and \eqref{eq:app_2timecorrs_evolution2}, we factorized the third-order correlations,
$\langle{\sigma}_{a}^z(t+\tau){\sigma}_{b}^+(t+\tau){\sigma}_{B}^-(t)\rangle
\approx
s_a^z({t+\tau}) \langle {\sigma}_{b}^+({t+\tau}){\sigma}_{B}^-(t)\rangle$.

\section{Trajectory calculations}
\label{app:trajectories}
We solve the stochastic master equation
\begin{equation}
\begin{split}
    \dot \rho_\meas =
    &-i[\frac{\Omega}{2}(a^\dag S^- + a S^+) + H_0 + H_\mathrm{inter},\rho_\meas] + 
    \\
    &+\cavdec \mathcal{D}[a_+]\rho_\meas + \mathcal{L}_\mathrm{drive}\rho_\meas +
    \\
    & + \frac{\mathrm{d}W}{\mathrm{d}t}
    \sqrt{\cavdec \xi}
    \left[e^{i\phi_\meas(t)} ( a\rho_\meas - \expval{a}_\meas\rho_\meas) + \mathrm{H.c.} \right]
    \end{split}
    \label{eq:app_master_stochastic}
\end{equation}
by employing a cumulant expansion to second order.
To this end, we compute equations of motions for expectation values of operators $a_+$, $a_+^\dag$, $\sigma_{A,B}^{z,\pm}$ and products of two such operators.
The equation of motion for each operator $o$ reads
$\mathrm{d}\expval{o}_\meas/\mathrm{d}t  = \Tr[\dot{\rho}_\meas o]$.
Any expectation value of a product of three operators is factorized using \cref{eq:app_cumulant}.
We invoke the permutational invariance to set all spins within each species equal, as was done in the finite-size calculations of the steady state, see \cref{app:finite_size_calculations}.
The noise explicitly breaks the $U(1)$-symmetry; therefore, we keep terms such as $\expval{\sigma^+_a}$ or $\expval{\sigma^z_a \sigma^+_b}$.

This leads to 27 equations, e.g.,  
\begin{equation}
\begin{split}
    \frac{\mathrm{d}}{\mathrm{d}t}\expval{a} = 
    &-\frac{\cavdec}{2} \expval{a} + iN\frac{\Omega}{2} (\expval{\sigma^+_A}+\expval{\sigma^+_B}) + 
    \\
    &+\sqrt{\cavdec\xi/2} \frac{\mathrm{d}W}{\mathrm{d}t}\Big[
    e^{i\phi_\meas(t)}(\expval{a a} - \expval{a}^2) + 
    \\
    &\phantom{+\sqrt{\cavdec\eta/2} \frac{\mathrm{d}W}{\mathrm{d}t}\Big[}
    e^{-i\phi_\meas(t)}(\expval*{a^\dag a} - \abs{\expval{a}}^2)
    \Big]
    \, .
\end{split}
\end{equation}
Note that the noise term vanishes in the thermodynamic limit when the mean-field factorization is exact.
We do not list all equations here; they can be derived and evaluated using the QuantumCumulants.jl package~\cite{Plankensteiner2022quantumcumulantsjl}.

\end{document}